\documentclass[envcountsect,envcountsame]{llncs}
\usepackage{amsmath,amssymb}
\usepackage{stmaryrd}
\usepackage[final]{graphicx}
\usepackage[draft=false]{hyperref}
\usepackage{cite}
\usepackage{wrapfig}
\usepackage{tikz}
\usepackage{latexsym}
\usepackage{caption}
\usetikzlibrary{automata,arrows,positioning}
\tikzset{
	initial text=$ $,
}

\usepackage{pgfplots}
\usetikzlibrary{positioning,shapes,shadows,arrows}
\usepackage[final]{listings}
\title{Tail Probabilities for  Randomized Program Runtimes via Martingales for Higher Moments}
\author{Satoshi Kura\inst{1,2} \and Natsuki Urabe\inst{1,2} \and Ichiro Hasuo\inst{2,3}}
\institute{Department of Computer Science, University of Tokyo,  Japan
\and National Institute of Informatics, Tokyo, Japan
\and The Graduate University for Advanced Studies (SOKENDAI), Kanagawa, Japan}

\newcommand{\upper}[1]{\overline{#1}}

\newcommand{\nexttime}{\mathbb{X}}
\newcommand{\real}{\mathbb{R}}
\newcommand{\transition}{\mapsto}
\newcommand{\elapse}[1]{\mathrm{El}_{#1}}
\newcommand{\reward}{\mathrm{Rew}}
\newcommand{\lfp}{\mu}
\newcommand{\prob}[1]{\mathrm{Pr}({#1})}
\newcommand{\expectation}[1]{\mathbb{E}[{#1}]}
\newcommand{\pushforward}[1]{({#1})_{*}}
\newcommand{\comp}{\mathrel{\circ}}

\newcommand{\Borel}{\mathcal{B}}
\newcommand{\moment}{\mathbb{M}}
\newcommand{\supp}{\mathrm{supp}}
\newcommand{\update}[3]{{#1}({#2} \leftarrow {#3})}
\newcommand{\configuration}{S}
\newcommand{\transpose}[1]{{#1}^{T}}
\newcommand{\init}{\mathrm{init}}
\newcommand{\run}[1]{\mathrm{Run}({#1})}
\newcommand{\linexpr}[2]{{#1}_{\mathrm{lin}}[{#2}]}
\newcommand{\polyexpr}[2]{{#1}[{#2}]}

\newcommand{\place}{\underline{\phantom{n}}\,}
\newcommand{\monomials}[1]{\mathcal{M}_{\leq #1}}
\renewcommand{\implies}{\,\Rightarrow\,}
\renewcommand{\paragraph}[1]{\vspace{1mm}\noindent{\bf #1}\;}

\newcounter{myfigure}

\newif\ifignore 
\ignoretrue 

\spnewtheorem{mytheorem}{Theorem}[section]{\bfseries}{\itshape} 
\spnewtheorem{mylemma}[mytheorem]{Lemma}{\bfseries}{\itshape}
\spnewtheorem{myproposition}[mytheorem]{Proposition}{\bfseries}{\itshape}
\spnewtheorem{mysublemma}[mytheorem]{Sublemma}{\bfseries}{\itshape}
\spnewtheorem{mycorollary}[mytheorem]{Corollary}{\bfseries}{\itshape}
\spnewtheorem{myfact}[mytheorem]{Fact}{\bfseries}{\itshape}
\spnewtheorem{mynotation}[mytheorem]{Notation}{\bfseries}{\rmfamily}
\spnewtheorem{myremark}[mytheorem]{Remark}{\bfseries}{\rmfamily}
\spnewtheorem{myexample}[mytheorem]{Example}{\bfseries}{\rmfamily}
\spnewtheorem{myassumption}[mytheorem]{Assumption}{\bfseries}{\rmfamily}
\spnewtheorem{mydefinition}[mytheorem]{Definition}{\bfseries}{\rmfamily}
\spnewtheorem{myrequirements}[mytheorem]{Requirements}{\bfseries}{\rmfamily}
\spnewtheorem{myproblem}[mytheorem]{Problem}{\bfseries}{\rmfamily}

\begin{document}
\maketitle

\begin{abstract}
Programs with randomization constructs is an active research topic,  especially after the recent introduction of martingale-based analysis methods for their termination and runtimes. Unlike most of the existing works that focus on proving almost-sure termination or estimating the expected runtime, in this work we study the \emph{tail probabilities} of runtimes---such as ``the execution takes more than 100 steps with probability at most 1\%.'' To this goal,  we devise a theory of supermartingales that overapproximate \emph{higher moments} of runtime. These higher moments, combined with a  suitable  concentration inequality, yield useful upper bounds of tail probabilities. Moreover, our vector-valued formulation enables automated  template-based synthesis of those supermartingales. Our experiments suggest the method's  practical use. 
\end{abstract}

\section{Introduction}
The important roles of \emph{randomization} in algorithms and software systems are nowadays well-recognized. In algorithms,  randomization can bring remarkable speed gain at the expense of small probabilities of imprecision. 
In cryptography, many encryption algorithms are randomized in order to conceal the identity of plaintexts.  In software systems, randomization is widely utilized for the purpose of fairness, security and privacy. 

Embracing randomization in programming languages has therefore been an active research topic for a long time. Doing so does not only offer a solid infrastructure that programmers and system designers can rely on, but also opens up the possibility of \emph{language-based, static} analysis of properties of randomized algorithms and systems. 

The current paper's goal is to analyze imperative programs with randomization constructs---the latter come in two forms, namely probabilistic branching and assignment from a designated, possibly continuous, distribution. We shall refer to such programs as \emph{randomized programs}.\footnote{With the  rise of statistical machine learning, \emph{probabilistic programs}   attract a lot of attention.
Randomized programs can be thought of as a fragment of probabilistic programs without \emph{conditioning} (or \emph{observation}) constructs. In other words, the Bayesian aspect of probabilistic programs is absent in randomized programs. }

\paragraph{Runtime and Termination Analysis of Randomized Programs}
The \emph{runtime} of a randomized program is often a problem of our interest; so is \emph{almost-sure termination}, that is, whether the program terminates with probability $1$. In the programming language community,  these problems have been taken up by many researchers as a challenge of both practical importance and theoretical interest.

Most of the existing works on runtime and termination analysis follow either of the following two approaches. 
\begin{itemize}
 \item \emph{Martingale-based methods}, initiated with a notion of \emph{ranking supermartingale} in~\cite{DBLP:conf/cav/ChakarovS13} and 
 extended~\cite{DBLP:journals/pacmpl/AgrawalC018,DBLP:conf/cav/ChatterjeeFG16,DBLP:conf/popl/FioritiH15,DBLP:journals/toplas/ChatterjeeFNH18,DBLP:conf/atva/JagtapSZ18}, have their origin in the theory of stochastic processes. They can also be seen as a 
 probabilistic extension of \emph{ranking functions}, a standard proof method for termination of (non-randomized) programs. Martingale-based methods have seen remarkable success in \emph{automated synthesis} using templates and constraint solving (like LP or SDP). 
 \item The \emph{predicate-transformer} approach,\! pursued in~\cite{DBLP:journals/jacm/KaminskiKMO18,DBLP:conf/esop/BatzKKM18,DBLP:conf/qest/KaminskiKM16},\! uses a more syntax-guided formalism of program logic and emphasizes reasoning by \emph{invariants}. 
\end{itemize}
The essential difference between the two approaches is not big: an invariant notion in the latter is easily seen to be an adaptation of a suitable notion of supermartingale. The work~\cite{DBLP:conf/atva/TakisakaOUH18} presents a comprehensive account on the order-theoretic foundation behind these techniques. 

 These  existing works are mostly focused on the following 
 problems:  deciding almost-sure termination, computing termination probabilities, and computing expected runtime. (Here ``computing'' includes giving upper/lower bounds.) See~\cite{DBLP:conf/atva/TakisakaOUH18} for a comparison of some of the existing martingale-based methods.

\paragraph{Our Problem: Tail Probabilities for Runtimes}
In this paper we focus on the problem of \emph{tail probabilities} that is not studied much so far.\footnote{An exception is~\cite{DBLP:journals/corr/ChatterjeeF17}; see~\S\ref{sec:related} for comparison with the current work.} We present a method for \emph{overapproximating} tail probabilities; here is the 
problem we solve. 
\begin{tabular}{rl}
 \textbf{Input:} & a randomized program $\Gamma$, and a \emph{deadline} $d\in\mathbb{N}$ \\
 \textbf{Output:} &an upper bound of the \emph{tail probability} $\prob{T_{\mathrm{run}} \geq d}$, where $T_{\mathrm{run}}$ is  \\
 & the runtime of $\Gamma$
\end{tabular}

Our target language is a imperative language that features randomization (probabilistic branching and random assignment). We also allow nondeterminism; this makes the program's runtime depend on the choice of a \emph{scheduler} (i.e.\ how nondeterminism is resolved). In this paper we study the longest, worst-case runtime (therefore our scheduler is \emph{demonic}). In the technical sections, we use the presentation of these programs as \emph{probabilistic control graphs (pCFGs)}---this is as usual in the literature. See e.g.~\cite{DBLP:journals/pacmpl/AgrawalC018,DBLP:conf/atva/TakisakaOUH18}. 

\begin{wrapfigure}[6]{r}{4.3cm}
\vspace{-1.cm}
\hspace*{2mm}
\scriptsize
\begin{minipage}{4.0cm}
\begin{lstlisting}[numbers=left,basicstyle={\scriptsize\ttfamily}]
x := 2;  y := 2;
while (x > 0 && y > 0) do
  z := Unif (-2,1);
  if * then
    x := x + z
  else
    y := y + z
  fi
od
\end{lstlisting}
\vspace{-2em}
\end{minipage}
\caption{An example program}\label{fig:exampleIntro}
\end{wrapfigure}
 An example of our target program is in Fig.~\ref{fig:exampleIntro}.
 It is an imperative program with randomization: in Line 3, the value of $z$ is sampled from the uniform distribution over the interval $[-2, 1]$. 
The symbol $\*$ in the line 4 stands for a nondeterministic Boolean value; in our analysis, it is resolved so that the runtime becomes the longest. 

Given the program in Fig.~\ref{fig:exampleIntro} and a choice of a deadline (say $d=400$), we can ask the question ``what is the  probability $\prob{T_{\mathrm{run}} \geq d}$ for the runtime $T_{\mathrm{run}}$ of the program to exceed $d=400$ steps?'' As we show in~\S{}\ref{sec:experiments}, our method gives a guaranteed upper bound $0.0684$. This means that, if we allow the time budget of $d=400$ steps, the program terminates  with the probability at least 93\%.

\begin{figure}[tbp]
 \begin{tikzpicture}[node distance=1.6mm and 5mm, every node/.style={inner sep=1pt}]
 \node (input) {a randomized program $\Gamma$};
 \node[draw, below=of input] (sm) {\textbf{step 1:} template-based synthesis of vector-valued supermartingales (\S\ref{sec:supermartingale}, \S\ref{sec:synthesis})};
 \node[below=of sm] (moment) {upper bounds of higher moments $\expectation{T_{\mathrm{run}}}, \dots, \expectation{(T_{\mathrm{run}})^K}$};
 \node[draw, below=of moment] (ineq) {\textbf{step 2:} calculation via a concentration inequality (\S\ref{sec:concentration_inequality})};
 \node[below=of ineq] (output) {an upper bound of the tail probability $\prob{T_{\mathrm{run}} \geq d}$};
 \draw[->, very thick] (input) -- (sm);
 \draw[->, very thick] (sm) -- (moment);
 \draw[->, very thick] (moment) -- (ineq);
 \draw[->, very thick] (ineq) -- (output);
 \node[left=of ineq] (dl) {a deadline $d$};
 \draw[->, very thick] (dl) -- (ineq);
 \end{tikzpicture}
 \vspace{-3mm}
\caption{Our workflow}\label{fig:ourWorkflow}
\vspace{-.8cm}
\end{figure}
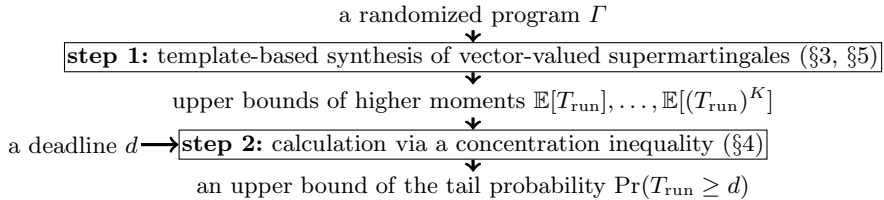

\paragraph{\fussy Our Method: Concentration Inequalities, Higher Moments, and\linebreak Vector-Valued Supermartingales}
Towards the goal of computing tail probabilities, our approach is to use \emph{concentration inequalities}, a  technique from probability theory that is commonly used for overapproximating various tail probabilities. There are various 
concentration inequalities in the literature, and each of them is applicable in a different setting, such as a nonnegative random variable (Markov's inequality), known mean and variance (Chebyshev's inequality), a difference-bounded martingale (Azuma's inequality), and so on. Some of them 
were
used for 
analyzing
randomized programs~\cite{DBLP:journals/corr/ChatterjeeF17} (see~\S{}\ref{sec:related} for comparison). 

In this paper, we use a specific concentration inequality that uses \emph{higher moments} $\expectation{T_{\mathrm{run}}}, \dots, \expectation{(T_{\mathrm{run}})^K}$ of  runtimes $T_{\mathrm{run}}$, up to a choice of the maximum degree $K$.  The concentration inequality is taken from~\cite{BLM2013Concentration}; it generalizes Markov's and Chebyshev's. We observe that a higher moment yields a tighter bound of the tail  probability, as the deadline $d$ grows bigger. Therefore it makes sense to strive for computing higher moments. 

For computing higher moments of runtimes, we systematically extend the existing theory of ranking supermartingales, from the expected runtime (i.e.\ the first moment) to higher moments. The theory features a \emph{vector-valued} supermartingale, which not only generalizes easily to degrees up to arbitrary $K\in\mathbb{N}$, but also allows automated synthesis much like usual supermartingales. 

We also claim that the soundness of these vector-valued supermartingales is proved in a mathematically clean manner. Following our previous work~\cite{DBLP:conf/atva/TakisakaOUH18}, our arguments are based on the order-theoretic foundation of fixed points (namely the Knaster-Tarski, Cousot--Cousot and Kleene theorems), and we give upper bounds of higher moments by suitable least fixed points. 

Overall,  our workflow is as shown in Fig.~\ref{fig:ourWorkflow}. We note that the step 2 in Fig.~\ref{fig:ourWorkflow} is computationally much cheaper than the step 1: in fact, the step 2 yields a symbolic expression for an upper bound in which $d$ is a free variable. This makes it possible to draw graphs like the ones in Fig.~\ref{fig:graph}. It is also easy to find a deadline $d$ for which $\prob{T_{\mathrm{run}} \geq d}$ is below a given threshold $p\in[0,1]$.

We implemented a prototype that synthesizes vector-valued supermartingales using linear and polynomial templates. The resulting constraints are solved  by LP and SDP solvers, respectively. 
 Experiments show that our method 
 can
 produce nontrivial upper bounds in reasonable computation time. 
 We also experimentally confirm that higher moments are useful in producing tighter bounds. 


\paragraph{Our Contributions}
Summarizing, the contribution of this paper is as follows.\vspace{-0.5\baselineskip}
\begin{itemize}
 \item We extend the existing theory of ranking supermartingales from expected runtimes (i.e.\ the first moment) to \emph{higher moments}. The extension has a solid foundation of order-theoretic fixed points. Moreover, its clean presentation by vector-valued supermartingales makes automated synthesis as easy as before. Our target randomized programs are rich, embracing nondeterminism and continuous distributions. 
 \item We study how 
these vector-valued supermartingales (and the resulting upper bounds of higher moments) can be used to yield upper bounds of \emph{tail probabilities of runtimes}.  We identify  a concentration lemma that suits  this purpose. We show that higher moments indeed yield tighter bounds.
 \item Overall, we present a comprehensive language-based framework for overapproximating tail probabilities of runtimes of randomized programs (Fig.~\ref{fig:ourWorkflow}). It has been implemented, and our experiments suggest its practical use. 
\end{itemize}\vspace{-0.6\baselineskip}

      

\paragraph{Organization} 
We give preliminaries in \S{}\ref{sec:preliminaries}.
In \S\ref{sec:supermartingale}, we review the order-theoretic characterization of ordinary ranking supermartingales and present an extension to higher moments of runtimes.
In \S\ref{sec:concentration_inequality}, we discuss how to obtain an upper bound of the tail probability of runtimes. 
In \S\ref{sec:synthesis}, we explain an automated synthesis algorithm for our ranking supermartingales.
In \S\ref{sec:experiments}, we give experimental results. 
In \S\ref{sec:related}, we discuss related work. We conclude and give future work in \S\ref{sec:conclusion}.
Some proofs and details are deferred to the appendices.

\vspace{-1.5\baselineskip}\mbox{}
\section{Preliminaries}\label{sec:preliminaries}\vspace{-0.5\baselineskip}

We present some preliminary materials, including the definition of pCFGs (we use them as a model of randomized programs) and the definition of runtime. 

Given topological spaces $X$ and $Y$, let $\Borel(X)$ be the set of Borel sets on $X$ and $\Borel(X, Y)$ be the set of Borel measurable functions $X \to Y$.
We assume that the set $\real$ of reals, a finite set $L$ and the set $[0, \infty]$ are equipped with the usual topology, the discrete topology, and the order topology, respectively. We use the induced Borel structures for these  spaces.
Given a measurable space $X$, let $\mathcal{D}(X)$ be the set of probability measures on $X$.
For any $\mu \in \mathcal{D}(X)$, let $\supp(\mu)$ be the support of $\mu$.
We write $\expectation{X}$ for the expectation of a random variable $X$.

Our use of pCFGs 
follows recent works including~\cite{DBLP:journals/pacmpl/AgrawalC018}.\vspace{-0.3\baselineskip}
\begin{mydefinition}[pCFG]\label{def:pCFG}
A \emph{probabilistic control flow graph (pCFG)} is a tuple $\Gamma = (L, V, l_{\init}, \vec{x}_{\init}, {\transition}, \mathrm{Up}, \mathrm{Pr}, G)$ that consists of the following.
\vspace{-0.5\baselineskip}
\begin{itemize}
\item 
A finite set $L$ of \emph{locations}.
It is a disjoint union of sets $L_D$, $L_P$, $L_n$ and $L_A$ of 
\emph{deterministic}, \emph{probabilistic}, \emph{nondeterministic} and \emph{assignment} locations.
%
\item A finite set $V$ of \emph{program variables}.
\item An \emph{initial location} $l_{\init} \in L$. 
\qquad\begin{minipage}{0.5\hsize}\renewcommand{\labelitemii}{$-$}\begin{itemize}\item An \emph{initial valuation} $\vec{x}_{\init} \in \real^{V}$\end{itemize}\end{minipage}
\item A \emph{transition relation} ${\transition} \subseteq L \times L$ which is total (i.e.\ $\forall l.\, \exists l'.\,l \transition l'$).
\item An \emph{update function} $\mathrm{Up} : L_A \to V \times \bigl(\,\Borel(\real^V, \real) \cup \mathcal{D}(\real) \cup \Borel(\real)\,\bigr)$ for assignment.
\item A family $\mathrm{Pr} = (\mathrm{Pr}_l)_{l \in L_P}$ of probability distributions, where $\mathrm{Pr}_l \in \mathcal{D}(L)$, for probabilistic locations. We require that
 $l' \in \supp(\mathrm{Pr}_l)$ implies $l \transition l'$. 
\item A \emph{guard function} $G : L_D \times L \to \Borel(\real^V)$ such that for each $l \in L_D$ and $\vec{x} \in \real^V$, there exists a unique location $l' \in L$ satisfying $l \transition l'$ and $\vec{x} \in G(l, l')$.
\end{itemize}\vspace{-0.4\baselineskip}
The update function  can be decomposed into three functions $\mathrm{Up}_D : L_{AD} \to V \times \Borel(\real^V, \real)$, $\mathrm{Up}_P : L_{AP} \to V \times \mathcal{D}(\real)$ and $\mathrm{Up}_N : L_{AN} \to V \times \Borel(\real)$, under a suitable decomposition $L_A = L_{AD} \cup L_{AP} \cup L_{AN}$ of assignment locations. The elements of
$L_{AD}$, $L_{AP} $ and $L_{AN}$ represent 
\emph{deterministic}, \emph{probabilistic} and \emph{nondeterministic} assignments, respectively.
See e.g.~\cite{DBLP:conf/atva/TakisakaOUH18}.
\end{mydefinition}

\begin{wrapfigure}[5]{r}{5.4cm}
%
\vspace{-2.8em}
\begin{tikzpicture}[scale=0.70, every node/.style={transform shape}, baseline=0, every state/.style={inner sep=2pt, minimum size=18pt}]
\node[state, initial above] (l0) at (-2.3, 0) {$l_0$};
\node[state] (l1) at (-1.15, 1.2) {$l_1$};
\node[state] (l2) at (0, -.1) {$l_2$};
\node[state] (l3) at (2.2, 0) {$l_3$};
\node[state, rectangle] (l4) at (4.7, 0) {$l_4$};
\node[state] (l5) at (2.5, 1.2) {$l_5$};
\node[state] (l6) at (2.5, -1.1) {$l_6$};
\node[state, accepting] (l7) at (-2.3, -1.0) {$l_7$};
\draw[->] (l0) -- (l1) node[midway, fill=white, inner sep=1pt] {$x := 2$};
\draw[->] (l1) -- (l2) node[midway, fill=white, inner sep=1pt] {$y := 2$};
\draw[->] (l2) -- (l3) node[midway, draw, fill=white, align=center, inner sep=2pt, pos=0.55] {$x > 0$\\and\\$y > 0$};
\draw[->] (l3) -- (l4) node[midway, above, inner sep=1pt] {$\begin{aligned}&z :=\\[-1pt]  &\mathrm{Unif}(-2, 1)\end{aligned}$};
\draw[->] (l4) to[bend right] (l5);
\draw[->] (l4) to[bend left] (l6);
\draw[->] (l5) to[bend right] node[midway, fill=white, inner sep=1pt] {$x := x + z$} (l2);
\draw[->] (l6) to[bend left] node[midway, fill=white, inner sep=1pt] {$y := y + z$} (l2);
\draw[->] (l2) -- (l7) node[midway, draw, fill=white, align=center, inner sep=2pt] {$x \leq 0$\\or\\$y \leq 0$};
\end{tikzpicture}
\end{wrapfigure}
An example of a pCFG is 
shown on the right. 
It models the program in Fig.~\ref{fig:exampleIntro}. The 
node $l_4$ is a nondeterministic location. $\mathrm{Unif}(-2, 1)$ is the uniform distribution on the interval $[-2, 1]$.

A \emph{configuration} of a pCFG $\Gamma$ is a pair $(l, \vec{x}) \in L \times \real^V$ of a location and a valuation.
We regard the set $\configuration = L \times \real^V$ of configurations is equipped with the product topology where $L$ is equipped with the discrete topology.
%
We say  a configuration $(l', \vec{x}')$  is a \emph{successor} of
 $(l, \vec{x})$, if $l\mapsto l'$ and the following hold.
\vspace{-0.4\baselineskip}
\begin{itemize}
\item If  $l \in L_D$, then $\vec{x}' = \vec{x}$ and $\vec{x} \in G(l, l')$.
\quad\begin{minipage}{0.4\hsize}\renewcommand{\labelitemii}{$-$}\begin{itemize}\item If  $l \in L_N \cup L_P$, then $\vec{x}' = \vec{x}$.\end{itemize}\end{minipage}
\item If  $l \in L_A$, then $\vec{x}' = \update{\vec{x}}{x_j}{a}$, where
 $\update{\vec{x}}{x_j}{a}$ denotes the vector obtained by replacing the $x_j$-component of $\vec{x}$ by $a$.
Here $x_{j}$ is such that
 $\mathrm{Up}(l) = (x_j, u)$, and $a$ is chosen as follows: 1)
$a = u(\vec{x})$ if $u \in \Borel(\real^V, \real)$; 2) 
 $a \in \supp(u)$ if $u \in \mathcal{D}(\real)$; and 3)
 $a \in u$ if $u \in \Borel(\real)$.
\end{itemize}\vspace{-0.4\baselineskip}
An \emph{invariant} of a pCFG $\Gamma$ is a measurable set $I \in \Borel(\configuration)$ such that $(l_{\init}, \vec{x}_{\init}) \in I$ and $I$ is closed under taking successors (i.e.\ if $c \in I$ and $c'$ is a successor of $c$ then $c' \in I$). Use of invariants is a common technique in automated synthesis of supermartingales~\cite{DBLP:journals/pacmpl/AgrawalC018}: it restricts configuration spaces and thus makes the constraints on supermartingales weaker.
It is also common to take an invariant as a measurable set~\cite{DBLP:journals/pacmpl/AgrawalC018}.
A \emph{run} of $\Gamma$ is an infinite sequence of configurations $c_0 c_1 \dots$ such that $c_0$ is the initial configuration $(l_{\init}, \vec{x}_{\init})$ and $c_{i+1}$ is a successor of $c_i$ for each $i$.
Let $\run{\Gamma}$ be the set of runs of $\Gamma$.

A \emph{scheduler} resolves nondeterminism: at a location in $L_{N}\cup L_{AN}$, it chooses a distribution of next configurations depending on the history of configurations visited so far.
Given a pCFG $\Gamma$ and a scheduler $\sigma$ of $\Gamma$, a probability measure $\nu_{\sigma}^{\Gamma}$ on $\run{\Gamma}$ is defined in the usual manner. 
See 
Appendix~\ref{appendix:DetailsandProofsfromsecPreliminaries} 
for details. 

\begin{mydefinition}[reaching time $T_{C}^{\Gamma}, T_{C,\sigma}^{\Gamma}$]\label{def:reachingTime}
 Let $\Gamma$ be a pCFG and $C\subseteq S$ be a set of configurations called a \emph{destination}. 
 The \emph{reaching time} to $C$ is a function $T_C^{\Gamma} : \run{\Gamma} \to [0, \infty]$ defined by
\begin{math}
 (T_C^{\Gamma})(c_{0}c_{1}\dotsc) = \mathop{\mathrm{argmin}}_{i\in\mathbb{N}} (c_{i}\in C)
\end{math}. 
 Fixing a scheduler $\sigma$ makes $T_{C}^{\Gamma}$ a random variable, since $\sigma$ determines
a probability measure $\nu_{\sigma}^{\Gamma}$ on $\run{\Gamma}$.  It is denoted by $T_{C,\sigma}^{\Gamma}$. 
\end{mydefinition}

Runtimes of pCFGs are a special case of reaching times, namely to the set of  terminating configurations.

The following higher moments are central to our framework. Recall that we are interested in demonic schedulers, i.e.\ those which make runtimes longer.
\begin{mydefinition}[$ \moment_{C, \sigma}^{\Gamma, k}$ and $\upper{\moment}_{C}^{\Gamma, k}$]
Assume the setting of Def.~\ref{def:reachingTime}, and let $k\in\mathbb{N}$ and $c \in \configuration$. We write $\moment_{C, \sigma}^{\Gamma, k}(c)$ for the $k$-th moment of the reaching time of $\Gamma$ from $c$ to $C$ under the scheduler $\sigma$, i.e.\ that is, 
\begin{math}
 \moment_{C, \sigma}^{\Gamma, k}(c) = \expectation{(T^{\Gamma_c}_{C,\sigma})^k} = \int (T^{\Gamma_c}_C)^k\, \mathrm{d} \nu_{\sigma}^{\Gamma_c}
\end{math}
where $\Gamma_c$ is a pCFG obtained from $\Gamma$ by changing the initial configuration to $c$.
Their supremum under varying $\sigma$  is denoted by
\begin{math}
 \upper{\moment}_{C}^{\Gamma, k} := \sup_{\sigma} \moment_{C, \sigma}^{\Gamma, k}
\end{math}.
\end{mydefinition}

\section{Ranking Supermartingale for Higher Moments}
\label{sec:supermartingale}
We introduce one of the main contributions in the paper, 
 a notion of ranking supermartingale that overapproximates higher moments. It is motivated by the following observation: martingale-based reasoning about the second moment must concur with one about the first moment. 
We conduct a systematic theoretical extension that features an order-theoretic foundation and vector-valued supermartingales. The theory accommodates nondeterminism and continuous distributions, too. 
We omit some details and proofs; they are in 
Appendix~\ref{appendix:DetailsandProofsfromsecsupermartingale}.



The fully general theory for higher moments will be presented in~\S{}\ref{subsec:generalHigherMoments}; we present its restriction to the second moments in~\S{}\ref{subsec:secondmoments} for readability. 

Prior to these, we review the existing theory of ranking supermartingales, through the lens of order-theoretic fixed points. In doing so we follow~\cite{DBLP:conf/atva/TakisakaOUH18}.




\begin{mydefinition}[``nexttime'' operation $\upper{\nexttime}$ (pre-expectation)]\label{def:preexpectation}
Given $\eta : \configuration \to [0, \infty]$, let $\upper{\nexttime} \eta : \configuration \to [0, \infty]$ be the function defined as follows.
\vspace{-0.5\baselineskip}
\begin{itemize}
\item If $l \in L_D$ and $\vec{x} \vDash G(l, l')$, then $(\upper{\nexttime} \eta) (l, \vec{x}) = \eta(l', \vec{x})$.
\item If $l \in L_P$, then $(\upper{\nexttime} \eta) (l, \vec{x}) = \sum_{l \transition l'} \mathrm{Pr}_l(l') \eta(l', \vec{x})$.
\item If $l \in L_N$, then $(\upper{\nexttime} \eta) (l, \vec{x}) = \sup_{l \transition l'} \eta(l', \vec{x})$.
\item If $l \in L_A$, $\mathrm{Up}(l) = (x_j, u)$ and $l \transition l'$,
if $u \in \Borel(\real^V, \real)$, then $(\upper{\nexttime} \eta) (l, \vec{x}) = \eta(l', \update{\vec{x}}{x_j}{u(\vec{x})})$;
if $u \in \mathcal{D}(\real)$, then $(\upper{\nexttime} \eta) (l, \vec{x}) = \int_{\real} \eta(l', \update{\vec{x}}{x_j}{y})\, \mathrm{d} u(y)$; and
if $u \in \Borel(\real)$, then $(\upper{\nexttime} \eta) (l, \vec{x}) = \sup_{y \in u} \eta(l', \update{\vec{x}}{x_j}{y})$.
\end{itemize}\vspace{-0.5\baselineskip}
\end{mydefinition}
Intuitively, $\upper{\nexttime} \eta$ is the expectation of $\eta$ after one transition. Nondeterminism is resolved by the maximal choice. 

We define $F_1 : (\configuration \to [0, \infty]) \to (\configuration \to [0, \infty])$ as follows. 
\[ 
\setlength{\abovedisplayskip}{4pt}
\setlength{\belowdisplayskip}{4pt}
(F_1 (\eta))(c) = \begin{cases}
1 + (\upper{\nexttime} \eta)(c) & c \in I \setminus C \\
0 & \text{otherwise}
\end{cases}
\quad
 \text{(Here ``$1+$'' accounts for time elapse)}
 \]
The function $F_1$ is an adaptation of the \emph{Bellman operator}, a classic notion in the theory of Markov processes.
A similar notion is used e.g.\ in~\cite{DBLP:journals/jacm/KaminskiKMO18}.
The function space $(S \to [0, \infty])$ is a complete lattice structure, because $[0, \infty]$ is; moreover 
$F_1$ is easily seen to be monotone. It is not hard to see either that
 the expected reaching time
$\upper{\moment}^{\Gamma, 1}_{C}$ 
to $C$ coincides with the least fixed point $\lfp F_1$.

The following theorem is fundamental in theoretical computer science. 
\begin{mytheorem}[Knaster--Tarski,~\cite{Tarski55}]
Let $(L, \leq)$ be a complete lattice and $f : L \to L$ be a monotone function.
The least fixed point $\mu f$ is the least prefixed point, i.e.\ 
$\lfp f = \min \{ l \in L \mid f(l) \leq l \}$\,.
\qed
\end{mytheorem}
The significance of the Knaster-Tarski theorem in verification lies in the induced proof rule: $f(l)\leq l\Rightarrow\mu f\leq l$. Instantiating to the expected reaching time $
\upper{\moment}^{\Gamma, 1}_{C}\!=\!
\mu F_{1}$, it means
$F_1 (\eta) \!\leq\! \eta\Rightarrow\upper{\moment}^{\Gamma, 1}_{C} \!\leq\! \eta$, i.e.\ an arbitrary prefixed point of $F_{1}$---which coincides with the notion of ranking supermartingale~\cite{DBLP:conf/cav/ChakarovS13}---overapproximates the expected reaching time. This proves soundness of ranking supermartingales. 



\subsection{Ranking Supermartingales for the Second Moments}\label{subsec:secondmoments}
We extend ranking supermartingales to the second moments. It paves the way to a fully general theory (up to the $K$-th moments) in~\S{}\ref{subsec:generalHigherMoments}.

The key in the martingale-based reasoning of expected reaching times 
(i.e.\ first moments) was that they are characterized as the least fixed point of 
a function $F_1$. 
Here it is crucial that for an arbitrary random variable $T$, we have $\expectation{T+1}=\expectation{T}+1$
and therefore we can calculate $\expectation{T+1}$ from $\expectation{T}$.
However, this is not the case for second moments. 
As $\expectation{(T+1)^2} = \expectation{T^2} + 2 \expectation{T} + 1$, 
calculating the second moment requires not only $\expectation{T^2}$ but also $\expectation{T}$.
This encourages us to define a vector-valued supermartingale.



\begin{mydefinition}[time-elapse function $\elapse{1}$]\label{def:timeElapseFunc}
A function $\elapse{1} \!:\! [0, \infty]^2 \to [0, \infty]^2$ is defined by
$\elapse{1}(x_1, x_2) = (x_1 + 1, x_2 + 2 x_1 + 1)$.
\end{mydefinition}

Then, an extension of $F_1$ for second moments can be defined as a combination of the time-elapse function $\elapse{1}$ and the pre-expectation $\upper{\nexttime}$.
\begin{mydefinition}[$F_2$]
Let $I$ be an invariant and $C \subseteq I$ be a Borel set.
We define $F_2 : (\configuration \to [0, \infty]^2) \to (\configuration \to [0, \infty]^2)$ by 
\[
\setlength{\abovedisplayskip}{4pt}
\setlength{\belowdisplayskip}{4pt}
(F_2 (\eta)) (c) = \begin{cases}
(\upper{\nexttime}(\elapse{1} \comp \eta))(c) & c \in I \setminus C \\
(0, 0) & \text{otherwise.} 
\end{cases} 
\]
Here $\upper{\nexttime}$ is applied componentwise: $(\upper{\nexttime} (\eta_1, \eta_2)) (c) = ((\upper{\nexttime} \eta_1) (c), (\upper{\nexttime} \eta_2) (c))$.
\end{mydefinition}

We can extend the complete lattice structure of $[0, \infty]$ 
to the function space $S \to [0, \infty]^2$ in a pointwise manner.
It is a routine to prove that $F_2$ is monotone with respect to this complete lattice structure. 
Hence $F_2$ has the least fixed point.
In fact, while $\upper{\moment}^{\Gamma, 1}_{C}$ was characterized as the least fixed point of $F_1$, 
a tuple $(\upper{\moment}^{\Gamma, 1}_{C}, \upper{\moment}^{\Gamma, 2}_{C})$ is \emph{not} the least fixed point of $F_2$ (cf.\ Example~\ref{ex:nondet} and Thm.~\ref{thm:second_moment_wo_nondet}).
However, the least fixed point of $F_2$ \emph{overapproximates} the tuple of moments.

\begin{mytheorem}
\label{thm:second_moment}
For any configuration $c \in I$,
$(\lfp F_2) (c) \ge (\upper{\moment}^{\Gamma, 1}_{C} (c), \upper{\moment}^{\Gamma, 2}_{C} (c))$\,.
\qed
\end{mytheorem}

Let $T^{\Gamma}_{C, \sigma, n} = \min\{ n, T^{\Gamma}_{C, \sigma} \}$.
To prove the above theorem, 
we inductively prove
\[
\setlength{\abovedisplayskip}{2pt}
\setlength{\belowdisplayskip}{2pt}
\textstyle(F_2)^n(\bot)(c) \ge \left(\int T^{\Gamma_c}_{C, \sigma, n}\, \mathrm{d} \nu_{\sigma}^{\Gamma_c}, \;\int (T^{\Gamma_c}_{C, \sigma, n})^2\, \mathrm{d} \nu_{\sigma}^{\Gamma_c} \right) \]
for each $\sigma$ and $n$, 
and take the supremum.
See
Appendix~\ref{appendix:DetailsandProofsfromsecsupermartingale} 
for more details.

Like ranking supermartingale for first moments, ranking supermartingale for second moments is defined as a prefixed point of $F_2$, 
i.e.\ a function $\eta$ such that $\eta \ge F_2 (\eta)$. However, we modify the definition for the sake of implementation.
\begin{mydefinition}[ranking supermartingale for second moments]\label{def:ranksup2nd}
A ranking supermartingale for second moments is a function $\eta : \configuration \to \real^2$ such that:
i) $\eta(c) \ge (\upper{\nexttime}(\elapse{1} \comp \eta))(c)$ for each $c \in I \setminus C$; and
ii) $\eta(c) \ge 0$ for each $c \in I$.
\end{mydefinition}
Here, the time-elapse function $\elapse{1}$ captures a positive decrease of the ranking supermartingale.
Even though we only have inequality in Thm.~\ref{thm:second_moment}, we can prove the following desired property of our supermartingale notion.

\begin{mytheorem}\label{thm:uppbound2nd}
If $\eta : \configuration \to \real^2$ is a supermartingale for second moments,
then $\bigl(\upper{\moment}^{\Gamma, 1}_{C}(c), \upper{\moment}^{\Gamma, 2}_{C}(c)\bigr) \le \eta(c)$ for each $c \in I$.
\qed
\end{mytheorem}

The following example and theorem show that we cannot replace $\geq$ with $=$ in Thm.~\ref{thm:second_moment} in general, 
but it is possible in the absence of nondeterminism.
%


\begin{wrapfigure}[5]{r}{3.8cm}
\vspace{-0.4cm}
\hspace{-4mm}
\begin{tikzpicture}[scale=0.6, every node/.style={transform shape}, every state/.style={inner sep=2pt, minimum size=20pt}]
\node[state, initial above] (l0) at (1, .4) {$l_0$};
%
%
\node[state] (l1) at (1.5, -.6) {$l_1$};
\node[state] (l2) at (2.0, .5) {$l_2$};
\node[state] (l3) at (3.0, 1.) {$l_3$};
\node[state] (l4) at (4.0, 1.) {$l_4$};
\node[state] (l5) at (5.0, 1.) {$l_5$};
\node[state] (l6) at (6.0, 1.) {$l_6$};
\node[state] (l7) at (7.0, 1.) {$l_7$};
\node[state] (l8) at (2.5, -.6) {$l_8$};
\node[state] (l9) at (3.5, -.6) {$l_9$};
\node[state] (l10) at (4.5, -.6) {$l_{10}$};
\node[state] (l11) at (5.5, -.6) {$l_{11}$};
\node[state, accepting] (l12) at (6.5, -.2) {$l_{12}$};
\draw[->] (l0) -- (l1);
\draw[->] (l1) -- (l2);
\draw[->] (l2) -- (l3) node[midway, above left] {$\displaystyle \frac{1}{2}$};
\draw[->] (l2) -- (l12) node[midway, fill=white, pos = 0.45, inner sep=2pt] {$\displaystyle \frac{1}{2}$};
\draw[->] (l3) -- (l4);
\draw[->] (l4) -- (l5);
\draw[->] (l5) -- (l6);
\draw[->] (l6) -- (l7);
\draw[->] (l7) -- (l12);
\draw[->] (l1) -- (l8);
\draw[->] (l8) -- (l9);
\draw[->] (l9) -- (l10);
\draw[->] (l10) -- (l11);
\draw[->] (l11) -- (l12);
\draw (l12) edge[loop right] (l12);
\end{tikzpicture}
\vspace{-2mm}
\end{wrapfigure}

\vspace{-5mm}\mbox{}
\begin{myexample}\label{ex:nondet}
The figure on the right shows a pCFG such that
$l_2 \in L_P$ and all the other locations are in $L_N$,
the initial location is $l_0$ and $l_{12}$ is a 
terminating location.
For the pCFG, the left-hand side of the inequality in Thm.~\ref{thm:second_moment} 
is $\lfp F_2 (l_0) = \left( 6, 37.5 \right)$.
In contrast,
if a scheduler $\sigma$ takes a transition from $l_1$ to $l_2$ with probability $p$, $(\moment^{\Gamma, 1}_{C, \sigma}(l_0), \moment^{\Gamma, 2}_{C, \sigma}(l_0)) = \left( 6 - \frac{1}{2} p, 36 - \frac{5}{2} p \right)$.
Hence
the right-hand side 
is $(\upper{\moment}^{\Gamma, 1}_{C}(l_0), \upper{\moment}^{\Gamma, 2}_{C}(l_0)) = (6, 36)$.
\end{myexample}

\begin{mytheorem}
\label{thm:second_moment_wo_nondet}
If $L_N = L_{AN} = \emptyset$, 
$\forall c\in I.\, (\lfp F_2) (c) = (\upper{\moment}^{\Gamma, 1}_{C}(c), \upper{\moment}^{\Gamma, 2}_{C}(c))$.
\qed
\end{mytheorem}

\subsection{Ranking Supermartingales for the Higher Moments}\label{subsec:generalHigherMoments}
We extend the result in~\S{}\ref{subsec:secondmoments}
 to  moments higher than second.

Firstly, the time-elapse function $\elapse{1}$ is generalized as follows. 
\begin{mydefinition}[time-elapse function $\elapse{1}^{K, k}$]\label{def:timeElapseFuncKth}
For  $K\!\in\!\mathbb{N}$ and $k\!\in\!\{1, \dots, K\}$, a function $\elapse{1}^{K, k} : [0, \infty]^K \to [0, \infty]$ is defined by $\elapse{1}^{K, k}(x_1, \dots, x_K) = 1 + \sum_{j=1}^k \binom{k}{j} x_j$. Here $\binom{k}{j}$ is the binomial coefficient. 
\end{mydefinition}

Again, a monotone function $F_K$ is defined as a combination of the time-elapse function $\elapse{1}^{K, k}$ and the pre-expectation $\upper{\nexttime}$.
\begin{mydefinition}[$F_K$]\label{def:FK}
Let $I$ be an invariant and $C \subseteq I$ be a Borel set.
We define $F_K : (\configuration \to [0, \infty]^K) \to (\configuration \to [0, \infty]^K)$ 
by $F_K(\eta)(c) = (F_{K, 1}(\eta)(c), \dots,$ $ F_{K, K}(\eta)(c))$, where
 $F_{K, k} : (\configuration \to [0, \infty]^K) \to (\configuration \to [0, \infty])$ is given by 
\[
\setlength{\abovedisplayskip}{3pt}
\setlength{\belowdisplayskip}{3pt}
(F_{K, k} (\eta)) (c) = \begin{cases}
(\upper{\nexttime}(\elapse{1}^{K, k} \comp \eta))(c) & c \in I \setminus C \\
0 & \text{otherwise.} \\
\end{cases} \]

%
\end{mydefinition}

As in Def.~\ref{def:ranksup2nd},
we define a 
supermartingale 
as a prefixed point of $F_K$.
\begin{mydefinition}[ranking supermartingale for $K$-th moments]\label{def:ranksupHigher}
We define $\eta_1, \dots, \eta_K : \configuration \to \real$ by $(\eta_1(c), \dots, \eta_K(c)) = \eta(c)$. 
A \emph{ranking supermartingale for $K$-th moments} is a function 
$\eta : \configuration \to \real^K$ such that for each $k$,
i) $\eta_k(c) \geq (\upper{\nexttime} (\elapse{1}^{K, k} \comp \eta_k))(c)$ for each $c \in I \setminus C$; and
ii) $\eta_k(c) \geq 0$ for each $c \in I$.
\end{mydefinition}

For higher moments, we can prove an analogous result to
Thm.~\ref{thm:uppbound2nd}.
\begin{mytheorem}\label{thm:uppboundHigher}
If $\eta$ is a supermartingale for $K$-th moments, then for each $c \in I$,
$(\upper{\moment}^{\Gamma, 1}_{C}(c), \dots, \upper{\moment}^{\Gamma, K}_{C}(c)) \leq \eta(c)$.
\qed
\end{mytheorem}

\section{From Moments to Tail Probabilities} 
\label{sec:concentration_inequality}
We 
discuss how to obtain upper bounds of tail probabilities of runtimes 
from upper bounds of higher moments of runtimes.
Combined with the result in~\S{}\ref{sec:supermartingale},
it induces a martingale-based method for overapproximating tail probabilities.

We use a concentration inequality. 
There are many choices of concentration inequalities 
(see e.g.~\cite{BLM2013Concentration}),
and
we use  
a variant of
Markov's inequality.
We prove that the concentration inequality is not only sound but also complete in a sense.


Formally, our goal is to calculate 
is an upper bound of $\prob{T^{\Gamma}_{C, \sigma} \ge d}$ for a given deadline $d > 0$, under the assumption that we know upper bounds $u_1, \dots, u_K$ of  moments $\expectation{T^{\Gamma}_{C, \sigma}}, \dots, \expectation{(T^{\Gamma}_{C, \sigma})^K}$. 
In other words, 
we want to overapproximate $\sup_{\mu} \mu([d, \infty])$ where $\mu$ ranges over the set of probability measures on $[0, \infty]$ satisfying 
$\left( \int x\, \mathrm{d}\mu(x), \dots, \int x^K\, \mathrm{d}\mu(x) \right) \le (u_1, \dots, u_K)$.

To answer this problem, we 
use a
generalized form of Markov's inequality.
\begin{myproposition}[{see e.g.~\cite[\S{}2.1]{BLM2013Concentration}}]
\label{prop:markov}
Let $X$ be a real-valued random variable and $\phi$ be a nondecreasing and nonnegative function.
For any $d \in \real$ with $\phi(d) > 0$,
\begin{equation*}
\setlength{\abovedisplayskip}{3pt}
\setlength{\belowdisplayskip}{3pt}
\prob{X \geq d} \leq \frac{\expectation{\phi(X)}}{\phi(d)} .
\end{equation*}
\qed
\end{myproposition}

By letting $\phi(x) = x^k$ in Prop~\ref{prop:markov}, we obtain the following inequality.
%
It gives an 
upper bound of the tail probability that is ``tight.'' 
\begin{myproposition}
\label{prop:tail_ineq}
Let $X$ be a nonnegative random variable.
Assume 
$\expectation{X^k} \leq u_k$ for each $k \in \{ 0, \dots, K \}$.
Then, for any $d > 0$,
\begin{equation}
\setlength{\abovedisplayskip}{3pt}
\setlength{\belowdisplayskip}{3pt}
\prob{X \ge d} \le \min_{0 \le k \le K} \frac{u_k}{d^k}.
\label{eq:tail_ineq}
\end{equation}
Moreover, this upper bound is tight: for any $d > 0$, there exists a probability measure such that the above equation holds.
\end{myproposition}
\begin{proof}
The former part is immediate from Prop~\ref{prop:markov}.
For the latter part, consider $\mu = p \delta_d + (1-p) \delta_0$ where $\delta_x$ is the Dirac measure at $x$ 
and $p$ is the value of the right-hand side of (\ref{eq:tail_ineq}).
\qed
\end{proof}

By combining Thm.~\ref{thm:uppboundHigher} 
with Prop.~\ref{prop:tail_ineq}, 
we obtain the following corollary. We can use it for overapproximating tail probabilities.
\begin{mycorollary}\label{cor:tailprobbound}
Let $\eta : \configuration \to \real^K$ be a ranking supermartingale for $K$-th moments. For each scheduler $\sigma$ and a deadline $d > 0$,
\begin{equation}
\setlength{\abovedisplayskip}{4pt}
\setlength{\belowdisplayskip}{4pt}
\prob{T^{\Gamma}_{C, \sigma} \ge d} \le \min_{0 \le k \le K} \frac{\eta_k(l_{\init}, \vec{x}_{\init})}{d^k}\,.
\label{eq:tail_ineq_sm}
\end{equation}
Here $\eta_0, \dots, \eta_K$ are 
defined by $\eta_0(c) = 1$ and $\eta(c) = (\eta_1(c), \dots, \eta_K(c))$. 
\qed
\end{mycorollary}
Note that if $K = 1$, Cor.~\ref{cor:tailprobbound} is essentially the same as~\cite[Thm. 4]{DBLP:journals/corr/ChatterjeeF17}.
Note also that for each $K$ there exists $d>0$ such that $\frac{\eta_K(l_{\init}, \vec{x}_{\init})}{d^K}=\min_{0 \le k \le K} \frac{\eta_k(l_{\init}, \vec{x}_{\init})}{d^k}$.
Hence higher moments become useful in overapproximating tail probabilities
as 
$d$ gets large. 
Later in~\S\ref{sec:experiments}, we demonstrate this fact experimentally. 


\section{Template-Based Synthesis Algorithm}\label{sec:synthesis}
We discuss an automated synthesis algorithm that calculates an upper bound for the $k$-th moment of the runtime of a pCFG using a supermartingale in Def.~\ref{def:ranksup2nd} or Def.~\ref{def:ranksupHigher}. 
It takes a pCFG $\Gamma$, an invariant $I$, a set $C \subseteq I$ of configurations, and a natural number $K$ as input and outputs an upper bound of $K$-th moment.

Our algorithm is adapted from existing template-based algorithms for synthesizing a ranking supermartingale (for first moments)
~\cite{DBLP:conf/cav/ChakarovS13,DBLP:journals/toplas/ChatterjeeFNH18,DBLP:conf/cav/ChatterjeeFG16}.
It fixes a linear or polynomial template with unknown coefficients for a supermartingale 
and using numerical methods like linear programming (LP) or semidefinite programming (SDP), calculate a valuation of the unknown coefficients so that the axioms of 
ranking supermartingale for $K$-th moments 
are satisfied.
%

We hereby briefly explain the algorithms.
See
Appendix~\ref{sec:synthesisApp} 
for details. 

\paragraph{Linear Template}
Our linear template-based algorithm
is adapted from~\cite{DBLP:conf/cav/ChakarovS13,DBLP:journals/toplas/ChatterjeeFNH18}.
We should assume that 
$\Gamma$, 
$I$ and 
$C$ 
are all ``linear'' in the sense that expressions appearing in $\Gamma$ are all linear and 
$I$ and $C$ are represented by linear inequalities.
To deal with assignments from a distribution like $x:=\mathrm{Norm}(0,1)$, 
we also assume that expected values of distributions appearing in $\Gamma$ are known.

The algorithm first fixes a template for a supermartingale:
for each location $l$, 
it fixes a $K$-tuple 
\(
\bigl(
\sum_{j=1}^{|V|} a^l_{j, 1} x_j + b^l_{1}, 
\ldots,
\sum_{j}^{|V|} a^l_{j, K} x_j + b^l_{K}
\bigr)
\)
of linear formulas. Here each $a^l_{j,i}$ and $b^l_i$ are unknown variables called \emph{parameters}.
The algorithm next collects conditions on the parameters so that the tuples constitute 
a ranking supermartingale for $K$-th moments. 
It results in a conjunction of formulas of a form
$\varphi_1\geq 0\wedge\cdots\wedge\varphi_m\geq 0\,\Rightarrow\, \psi\geq 0$.
Here $\varphi_1,\ldots,\varphi_m$ are linear formulas without parameters and
$\psi$ is a linear formula where parameters linearly appear in the coefficients.
By  
Farkas' lemma (see e.g.~{\cite[Cor.~7.1h]{Schrijver86TLIP}})
we can turn such formulas into linear inequalities over parameters by adding new variables.
Its feasibility is efficiently solvable with an LP solver.
We naturally wish to minimize an upper bound of the $K$-th moment, 
i.e.\ the last component of $\eta(l_{\init}, \vec{x}_{\init})$. 
We can minimize it by setting it to the objective function of the LP problem. 

\paragraph{Polynomial Template}
The polynomial template-based algorithm is based on~\cite{DBLP:conf/cav/ChatterjeeFG16}.
This time, 
$\Gamma$, $I$ and $C$ 
can be ``polynomial.''
To deal with assignments of distributions,
we assume that the $n$-th moments of distributions 
in $\Gamma$ are easily calculated for each $n\in\mathbb{N}$.
It is similar to the linear template-based one. 

It first fixes a polynomial template for a supermartingale, i.e.\ it 
assigns each location $l$ a $K$-tuple of polynomial expressions with unknown coefficients. 
Likewise the linear template-based algorithm, the algorithm reduces the axioms of supermartingale for higher moments
to a conjunction of formulas of a form $\varphi_1\geq 0\wedge\cdots\wedge\varphi_m\geq 0\implies \psi\geq 0$.
This time, each $\varphi_i$ is a polynomial formula without parameters and $\psi$ is a polynomial formula whose coefficients are 
\emph{linear} formula over the parameters.
In the polynomial case, a conjunction of such formula is reduced to an SDP problem using a theorem called
Positivstellensatz (we used a variant called Schm\"udgen's Positivstellensatz~\cite{Schmudgen1991}). 
We solve the resulting problem using an SDP solver setting $\eta(l_{\init}, \vec{x}_{\init})$ as the objective function.

\section{Experiments}
\label{sec:experiments}
%
We implemented two programs in OCaml to synthesize a supermartingale based on a) a linear template and b) a polynomial template. 
The programs translate a given randomized program to a pCFG and output an LP 
or SDP problem as described in \S\ref{sec:synthesis}.
An invariant $I$ and a terminal configuration $C$ for the input program are specified manually.
See e.g.~\cite{KatoenMMM10} for automatic synthesis of an invariant.
For linear templates, we have used GLPK (v4.65)~\cite{glpk} as an LP solver.
For polynomial templates, we have used SOSTOOLS (v3.03)~\cite{sostools} (a sums of squares optimization tool that internally uses an SDP solver) on Matlab (R2018b).
We used SDPT3 (v4.0)~\cite{sdpt3} as an SDP solver. 
The experiments were carried out on a Surface Pro 4 with an Intel Core i5-6300U (2.40GHz) and 8GB RAM.
We tested our implementation for the following two programs and their variants,
which were also used in the literature~\cite{DBLP:journals/jacm/KaminskiKMO18,DBLP:journals/toplas/ChatterjeeFNH18}.
Their code is in Appendix~\ref{sec:testprogs}.

\noindent\textit{Coupon collector's problem.} 
A probabilistic model of collecting coupons enclosed in cereal boxes.
There exist $n$ types of coupons, and 
one repeatedly buy cereal boxes until all the types of coupons are collected. 
We consider two cases: (1-1) $n=2$ and (1-2) $n=4$.
We tested the linear template program for them.


\sloppy
\noindent\textit{Random walk.}
We  used three variants of 1-dimensional random walks: (2-1) integer-valued one, (2-2) real-valued one with assignments from continuous distributions, (2-3) with adversarial nondeterminism; and two variants of 2-dimensional random walks (2-4) and (2-5) with assignments from continuous distributions and adversarial nondeterminism.
We tested both the linear 
and the polynomial template programs for these examples.

\fussy

\paragraph{Experimental results}
We measured execution times needed for Step 1 in Fig.~\ref{fig:ourWorkflow}.
The results are in Table~\ref{tab:result}.
Execution times are less than 0.2 seconds for linear template programs and several minutes for polynomial template programs.
Upper bounds of tail probabilities obtained from Prop.~\ref{prop:tail_ineq} are in Fig.~\ref{fig:graph}.

We can see that
our method is applicable even with nondeterministic branching ((2-3), (2-4) and (2-5)) or assignments from continuous distributions ((2-2), (2-4) and (2-5)).
We can use a linear template 
for bounding 
higher moments as long as there exists a supermartingale for higher moments representable by linear expressions ((1-1), (1-2) and (2-3)).
In contrast, for (2-1), (2-2) and (2-4), only a polynomial template program found a supermartingale for second moments.

It is expectable that the polynomial template program gives a better bound than the linear one 
because a polynomial template is more expressive than a linear one.
However, it did not hold for some test cases,
probably because of numerical errors of the SDP 
solver.
For example, (2-1) has a supermartingale for third moments that can be checked by a hand calculation, but the SDP solver returned ``infeasible'' in the polynomial template program.
It appears that 
our program
fails when large numbers are involved (e.g.\ the third moments of (2-1), (2-2) and (2-3)).
We have also tested a variant of (2-1) where the initial position is multiplied by 10000.
Then the SDP solver returned ``infeasible'' in the polynomial template program
while the linear template program returns a nontrivial bound.
Hence it seems that numerical errors are likely to occur to the polynomial template program when large numbers are involved. 


\begin{table}[p]
\noindent
\begin{minipage}{0.6\hsize}
\centering 
\scalebox{.72}{\begin{tabular}{|c|c||c|c|c|c|c|}
\hline
& & \multicolumn{2}{c|}{a) linear template} & \multicolumn{3}{c|}{b) polynominal template} \\
\cline{2-7}
& moment & upper bound & time (sec) & upper bound & time (sec) & degree \\
\hline
& 1st & 13 & 0.012 & & & \\
\cline{2-7}
(1-1) & 2nd & 201 & 0.019 & & & \\
\cline{2-7}
& 3rd & 3829 & 0.023 & & & \\
\hline
& 1st & 68 & 0.024 & & & \\
\cline{2-7}
& 2nd & 3124 & 0.054 & & & \\
\cline{2-7}
(1-2) & 3rd & 171932 & 0.089 & & & \\
\cline{2-7}
& 4th & 12049876 & 0.126 & & & \\
\cline{2-7}
& 5th & 1048131068 & 0.191 & & & \\
\hline
& 1st & 20 & 0.024 & 20.0 & 24.980 & 2 \\
\cline{2-7}
(2-1) & 2nd & - & 0.013 & 2320.0 & 37.609 & 2 \\
\cline{2-7}
& 3rd & - & 0.017 & - & 30.932 & 3 \\
\hline
& 1st & 75 & 0.009 & 75.0 & 33.372 & 2 \\
\cline{2-7}
(2-2) & 2nd & - & 0.014 & 8375.0 & 73.514 & 2 \\
\cline{2-7}
& 3rd & - & 0.021 & - & 170.416 & 3 \\
\hline
& 1st & 62 & 0.020 & 62.0 & 40.746 & 2 \\
\cline{2-7}
(2-3) & 2nd & 28605.4 & 0.038 & 6710.0 & 97.156 & 2 \\
\cline{2-7}
& 3rd & 19567043.36 & 0.057 & - & 35.427 & 3 \\
\hline
(2-4) & 1st & 96 & 0.020 & 95.95 & 157.748 & 2 \\
\cline{2-7}
& 2nd & - & 0.029 & 10944.0 & 361.957 & 2 \\
\cline{2-7}
\hline
(2-5) & 1st & 90 & 0.022 & - & 143.055 & 2 \\
\cline{2-7}
& 2nd & - & 0.042 & - & 327.202 & 2 \\
\cline{2-7}
\hline
\end{tabular}}
\vspace{3mm}
\caption{Upper bounds of the moments of runtimes.
``-'' indicates that the LP or SDP solver returned ``infeasible''.
The ``degree'' column shows the degree of the polynomial template used in the experiments.}
\label{tab:result}
\end{minipage}
\hspace{3mm}
\begin{minipage}{0.32\hsize}
\hspace*{5mm}
\begin{minipage}{0.85\hsize}
\begin{lstlisting}[numbers=left,basicstyle={\ttfamily\scriptsize},numbersep=7pt]
x := 200000000;
while true do
  if prob(0.7) then
    z := Unif(0,1);
    x := x - z
  else
    z := Unif(0,1);
    x := x + z
  fi;
  refute (x < 0)
od
\end{lstlisting}
\end{minipage}
\vspace{-0.cm}
\setcounter{myfigure}{\thefigure}
\stepcounter{figure}
\captionof{figure}{A variant of (2-2).}\label{fig:very_long_runtime}
\setcounter{figure}{\themyfigure}
\end{minipage}
%
\centering
\includegraphics[scale=0.63]{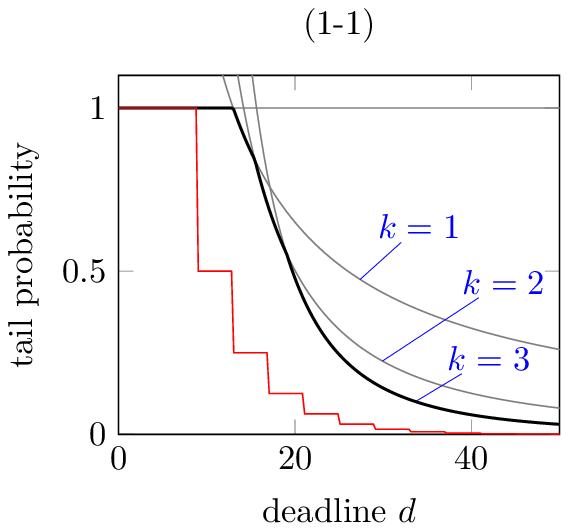}
\includegraphics[scale=0.63]{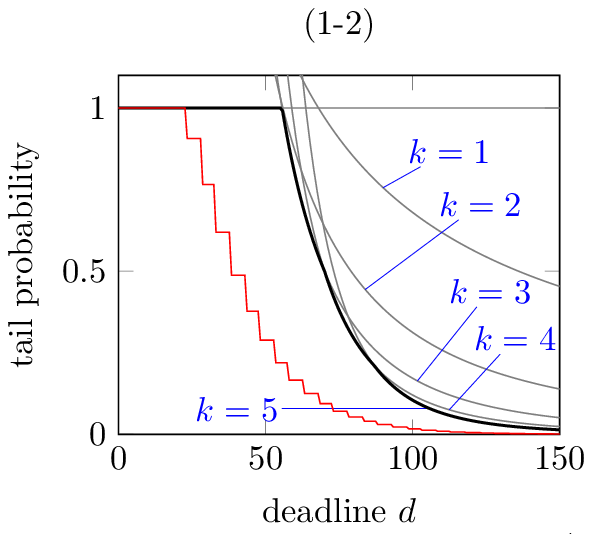}
\includegraphics[scale=0.63]{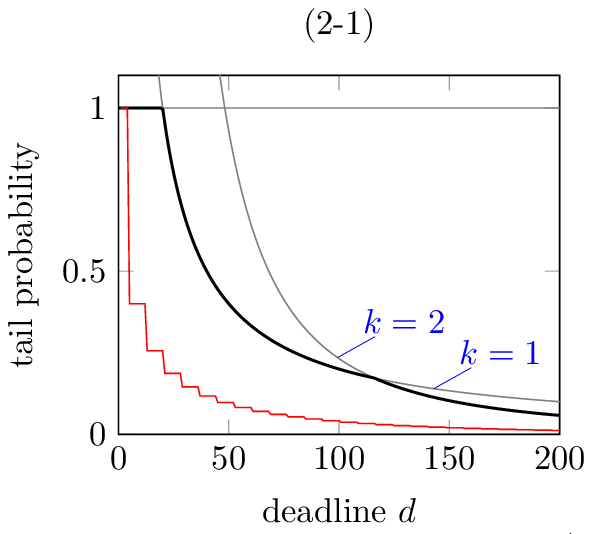}

\includegraphics[scale=0.63]{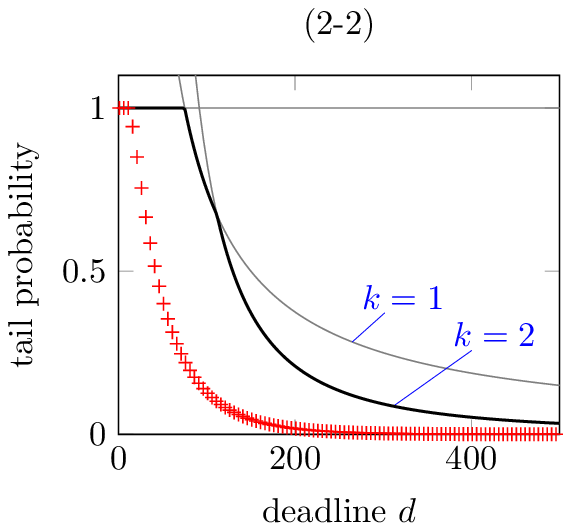}
\includegraphics[scale=0.63]{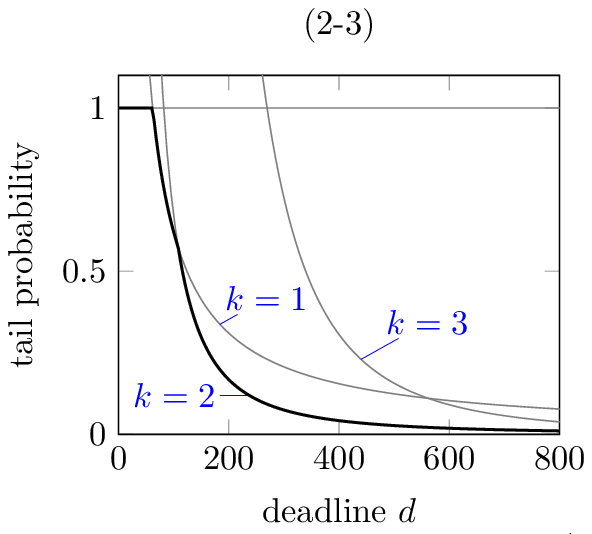}
\includegraphics[scale=0.63]{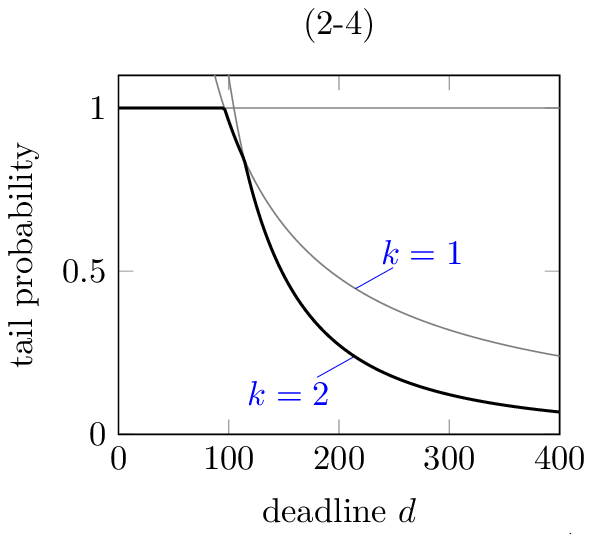}
\captionof{figure}{Upper bounds of the tail probabilities (except (2-5)).
Each gray line is the value of $\frac{u_k}{d^k}$ where $u_k$ is the best upper bound in Table~\ref{tab:result} of $k$-th moments and $d$ is a deadline.
Each black line is the minimum of gray lines, i.e.\ the upper bound by Prop.~\ref{prop:tail_ineq}.
The red lines in (1-1), (1-2) and (2-1) show the true tail probabilities calculated analytically.
The red points in (2-2) show tail probabilities calculated by Monte Carlo sampling where the number of trials is 100000000.
We did not calculate the true tail probabilities nor approximate them for (2-4) and (2-5) because these examples seem difficult to do so due to nondeterminism.}\label{fig:graph}
\stepcounter{figure}
\end{table}

Fig.~\ref{fig:graph} shows that the bigger the deadline $d$ is, the more useful higher moments become (cf.\ a remark just after Cor.~\ref{cor:tailprobbound}).
For example, in (1-2), an upper bound of $\prob{T^{\Gamma}_{C, \sigma} \geq 100}$ calculated from the upper bound of the first moment is 
$0.680$ while 
that of the fifth moment is $0.105$.

To show the merit of our method compared with sampling-based methods, we 
calculated a tail probability bound for
 a variant of (2-2) (shown in Fig.~\ref{fig:very_long_runtime} on p.~\pageref{fig:very_long_runtime})) with a deadline $d = 10^{11}$. %
Because of its very long expected runtime, 
a sampling-based method would not work for it.
In contrast, the linear template-based program gave an upper bound $\prob{T^{\Gamma}_{C, \sigma} \ge 10^{11}} \le 5000000025 / 10^{11} \approx 0.05$ in almost the same execution time as 
(2-2) ($<0.02$ seconds).

\section{Related Work}\label{sec:related}

\paragraph{Martingale-Based Analysis of Randomized Programs}
        Martingale-based methods are widely studied for the termination  analysis of randomized programs. One of the first 
        is \emph{ranking supermartingales}, introduced in~\cite{DBLP:conf/cav/ChakarovS13} for proving almost sure termination. The theory of ranking supermartingales has since been extended actively: accommodating nondeterminism~\cite{DBLP:journals/pacmpl/AgrawalC018,DBLP:conf/cav/ChatterjeeFG16,DBLP:conf/popl/FioritiH15,DBLP:journals/toplas/ChatterjeeFNH18}, syntax-oriented composition of supermartingales~\cite{DBLP:conf/popl/FioritiH15}, proving properties beyond termination/reachability~\cite{DBLP:conf/atva/JagtapSZ18}, and so on. Automated template-based synthesis of supermartingales by constraint solving 
        has been pursued, too~\cite{DBLP:conf/cav/ChakarovS13,DBLP:journals/toplas/ChatterjeeFNH18,DBLP:journals/pacmpl/AgrawalC018,DBLP:conf/cav/ChatterjeeFG16}.

Other martingale-based methods that are fundamentally different from ranking supermartingales have been devised, too. They include: different notions of \emph{repulsing supermartingales} for refuting termination (in~\cite{DBLP:conf/popl/ChatterjeeNZ17,DBLP:conf/atva/TakisakaOUH18}; also studied in control theory~\cite{steinhardtT12IJRR}); and \emph{multiply-scaled submartingales} for underapproximating reachability probabilities~\cite{urabeHH17LICS,DBLP:conf/atva/TakisakaOUH18}. See~\cite{DBLP:conf/atva/TakisakaOUH18} for an overview. 

In the literature on martingale-based methods, the one closest to this work is~\cite{DBLP:journals/corr/ChatterjeeF17}. Among its contribution is the analysis of tail probabilities. It is done by either of the following combinations: 1) \emph{difference-bounded} ranking supermartingales and the corresponding Azuma's 
concentration inequality; and 2) (not necessarily difference-bounded) ranking supermartingales and Markov's concentration inequality. When we compare these two methods with ours, the first method requires repeated martingale synthesis for different parameter values, which can pose a performance challenge. The second method corresponds to the restriction of our method to the first moment; recall that we showed the advantage of using higher moments, theoretically (\S\ref{sec:concentration_inequality}) and experimentally (\S{}\ref{sec:experiments}). 
See
Appendix~\ref{appendix:comparisonWithChatterjeeF17} 
for detailed discussions. Implementation is lacking in~\cite{DBLP:journals/corr/ChatterjeeF17}, too. 

We use Markov's inequality to calculate an upper bound of $\prob{T_{\mathrm{run}} \geq d}$ from a ranking supermartingale.
In~\cite{DBLP:journals/toplas/ChatterjeeFNH18}, Hoeffding's and Bernstein's inequalities are used for the same purpose. 
As the upper bounds obtained by these inequalities are exponentially decreasing with respect to $d$, 
they are asymptotically tighter than our bound obtained by Markov's inequality, assuming that we use the same ranking supermartingale. 
However, Hoeffding's and Bernstein's inequalities are applicable to limited classes of ranking supermartingales
(so-called difference-bounded and incremental ones, respectively). There exists a randomized program whose tail probability for runtimes is decreasing only polynomially 
(not exponentially, see 
Appendix~\ref{appendix:exTailProb}%
); this witnesses that there are cases where the methods in~\cite{DBLP:journals/toplas/ChatterjeeFNH18} do not apply but ours can.

 

The work~\cite{DBLP:journals/pacmpl/AgrawalC018} is also close to ours in that their supermartingales are vector-valued. The difference is in the orders: in~\cite{DBLP:journals/pacmpl/AgrawalC018} they use the \emph{lexicographic} order between vectors, and they aim 
to prove almost sure termination. In contrast, we use the \emph{pointwise} order between vectors, for 
overapproximating higher moments.

\paragraph{The Predicate-Transformer Approach to Runtime Analysis} 
In the runtime/termination analysis of randomized programs, another principal line of work uses \emph{predicate transformers}~\cite{DBLP:journals/jacm/KaminskiKMO18,DBLP:conf/esop/BatzKKM18,DBLP:conf/qest/KaminskiKM16}, following the precedent works on probabilistic predicate transformers such as~\cite{MorganMS96,Kozen81}. In fact, from the mathematical point of view, the  main construct for witnessing runtime/termination in those predicate transformer calculi (called \emph{invariants}, see e.g.\ in~\cite{DBLP:journals/jacm/KaminskiKMO18}) is essentially the same thing as ranking supermartingales. Therefore the difference between the martingale-based and predicate-transformer approaches is mostly the matter of presentation---the predicate-transformer approach is more closely tied to program syntax and has a stronger deductive flavor. It also seems that there is less work on automated synthesis in the predicate-transformer approach. 

In the predicate-transformer approach, the work~\cite{DBLP:conf/qest/KaminskiKM16}
 is the closest to ours, in that it studies \emph{variance} of runtimes of randomized programs. The main differences are as follows: 1) computing tail probabilities is not pursued~\cite{DBLP:conf/qest/KaminskiKM16}; 2) their extension from expected runtimes to variance involves an additional variable $\tau$, which poses a challenge in automated synthesis as well as in generalization to even higher moments; and 3) they do not pursue automated analysis.  See Appendix~\ref{appendix:comparisonWithKaminskiKM16} for further details.

\paragraph{Higher Moments of Runtimes}
Computing and using higher moments of runtimes of probabilistic systems---generalizing randomized programs---has been pursued before. 
In~\cite{DBLP:journals/siammax/DayarA05},
computing moments of runtimes of \emph{finite-state} Markov chains is reduced to a certain linear equation. In the study of randomized algorithms, the survey~\cite{doerr18probtools} collects a number of methods, among which are some tail probability bounds using higher moments. Unlike ours, none of these methods are language-based static ones. They do not allow automated analysis. 



\paragraph{Other Potential Approaches to Tail Probabilities}
We discuss potential approaches to estimating tail probabilities, other than the martingale-based one. 

\emph{Sampling} is widely employed for approximating behaviors of probabilistic systems;  especially so in the field of probabilistic programming languages, since exact symbolic reasoning is hard in presence of conditioning. See e.g.~\cite{DBLP:conf/ifl/TolpinMYW16}. We also used sampling to estimate tail probabilities in (2-2), Fig.~\ref{fig:graph}. The main advantages of our current approach over sampling are threefold: 1) our upper bounds come with a mathematical guarantee, while the sampling bounds can always be erroneous; 2) it requires ingenuity to sample programs with nondeterminism; and 3)  programs whose execution can take millions of years can still be analyzed by our method in a reasonable time, without executing them. The latter advantage is shared by static, language-based analysis methods in general; see e.g.~\cite{DBLP:conf/esop/BatzKKM18}. 

Another potential method is  probabilistic model checkers such as PRISM~\cite{KwiatkowskaNP11}. 
 Their algorithms are usually only applicable to finite-state models, and thus not to randomized programs in general. Nevertheless, fixing a deadline $d$ can make the reachable part $S_{\le d}$ of the configuration space $S$ finite, opening up the possibility of use of model checkers. It is an open question how to do so precisely, and  the following challenges are foreseen: 1) if the program contains continuous distributions,  the reachable part $S_{\le d}$ becomes infinite; 2) even if $S_{\le d}$ is finite, one has to repeat  (supposedly expensive) runs of a model checker for each choice of $d$. In contrast, in our method, an upper bound for the tail probability $\prob{T_{\mathrm{run}} \geq d}$ is symbolically expressed as a function of $d$ (Prop.~\ref{prop:tail_ineq}). Therefore, estimating tail probabilities for varying $d$ is computationally cheap. 

\section{Conclusions and Future Work}\label{sec:conclusion}
We provided a technique to obtain an upper bound of the tail probability of runtimes given a randomized algorithm and a deadline.
We first extended the ordinary ranking supermartingale notion 
using the order-theoretic characterization so that
it can 
calculate 
upper bounds of higher moments of runtimes for randomized programs.
Then 
by using a suitable concentration inequality,
we introduced
a method to calculate an upper bound of tail probabilities from upper bounds of higher moments. Our method is not only sound but also complete in a sense.
Our method was obtained by combining our supermartingale and the concentration inequality.
We also implemented an automated synthesis algorithm and demonstrated the applicability of our framework.

\paragraph{Future Work}
%
Example~\ref{ex:nondet} shows that our supermartingale is not complete: 
it sometimes fails to give a tight bound for higher moments.
Studying and improving the incompleteness is one possible direction of 
future work. 
For example, 
the following questions would be interesting:
Can bounds given by our supermartingale be arbitrarily bad?
Can we remedy the completeness by restricting the type of nondeterminism?
Can we define a complete supermartingale? 

Making our current method compositional is another direction of future research. Use of continuations, as in~\cite{DBLP:conf/esop/KaminskiKMO16}, can be a technical solution.
%

We are also interested in improving the implementation.
The polynomial template program failed to give an upper bound for higher moments because of numerical errors (see \S{}\ref{sec:experiments}).
We wish to remedy this situation. 
There exist several studies for using numerical solvers for verification without affected by numerical errors~\cite{jansson05TVI,janssonCK07REB,Jansson06VSDPTP,RouxIC18TACAS,RouxVS18FMSD}.
We might make use of these works for improvements.

\paragraph{Acknowledgement.}
We thank the anonymous referees for useful comments.
The authors are supported by JST ERATO HASUO Metamathematics for Systems Design Project (No.\ JPMJER1603), the JSPS-INRIA Bilateral Joint Research Project ``CRECOGI,'' and JSPS KAKENHI Grant No.\ 15KT0012 \& 15K11984. Natsuki Urabe is supported by JSPS KAKENHI Grant No.\ 16J08157.

\bibliographystyle{plain}
\bibliography{ref}

%

\clearpage
\appendix
\noindent{\Large\bf Appendix}

\section{Preliminaries on Measure Theory}\label{appendix:prelimMT}
In this section, we review some results from measure theory that is needed in the rest of the paper.
For more details, see e.g.\ \cite{AshDole99,tao2011introduction}.

\begin{mydefinition}
Let $\phi : X \to Y$ be a measurable function and $\mu$ be a probability measure on $X$.
A \emph{pushforward measure} $\pushforward{\phi} \mu$ is a measure on $Y$ defined by
$\pushforward{\phi} \mu(E) = \mu(\phi^{-1}(E))$
for each measurable set $E \subseteq Y$.
\end{mydefinition}

\begin{mylemma}\label{lem:pushforward}
Let $\phi : X \to Y$ and $f : Y \to [0, \infty]$ be measurable functions and $\mu$ be a probability measure on $X$.
\begin{equation}
\int f\, \mathrm{d}\big( \pushforward{\phi} \mu \big) = \int (f \comp \phi)\, \mathrm{d}\mu
\end{equation}
where $f \comp \phi$ denotes the composite function of $f$ and $\phi$.
\qed
\end{mylemma}

\begin{mylemma}\label{lem:Giry_Kleisli}
Let $(X, B_X)$ and $(Y, B_Y)$ be measurable spaces and $\mu_x$ be a probability measure on $Y$ for each $x \in X$.
The following conditions are equivalent.
\begin{enumerate}
\item \label{enum:ev1} For each $E \in B_Y$, a mapping $x \mapsto \mu_x(E)$ is measurable.
\item \label{enum:ev2} For each measurable function $f : X \times Y \to [0, \infty]$,
\[ x \mapsto \int_Y f(x, y) d\mu_x(y) \]
is measurable.
\end{enumerate}
\end{mylemma}
\begin{proof}
\paragraph{(\ref{enum:ev1} $\implies$ \ref{enum:ev2})}
We write $B_{X \times Y}$ for the product $\sigma$-algebra of $B_X$ and $B_Y$.
By the monotone convergence theorem (see e.g.\ \cite[Theorem~1.6.2]{AshDole99}) and the linearity of integration, it suffices to prove that for each $E \in B_{X \times Y}$, $f = 1_E : X \times Y \to [0, \infty]$ satisfies the condition \ref{enum:ev2}.
Let $M = \{ E \in B_{X \times Y} \mid \text{$1_E$ satisfies the condition \ref{enum:ev2}} \}$.
By the monotone class theorem (see e.g.\ \cite[Theorem~1.3.9]{AshDole99}), to prove $M = B_{X \times Y}$, it suffices to prove that $M$ is a monotone class and contains a Boolean algebra
\[ \{ \bigcup_{i=1}^n (E_i \times F_i) \mid E_i \in B_X, F_i \in B_Y \}. \]
The rest of the proof is easy.

\paragraph{(\ref{enum:ev2} $\implies$ \ref{enum:ev1})}
Given $E \in B_Y$, consider $f = 1_{X \times E}$.
\qed
\end{proof}

For any $f : A \to B$ and any set $X$, $X^f : X^B \to X^A$ denotes a precomposition of $f$ i.e.\ $X^f(u) = u \comp f$.
If $A \subseteq B$, we write $X^{A \subseteq B}$ for $X^i : X^B \to X^A$ where $i : A \to B$ is the inclusion mapping.

In \S\ref{appendix:DetailsandProofsfromsecPreliminaries}, we use the following corollary of the Kolmogorov extension theorem (see \cite[\S 2.4]{tao2011introduction}).
\begin{mycorollary}\label{cor:Kolmogorov}
Let $(X, B_X)$ be a measurable space and $\mu_n$ be an inner regular probability measure on $X^n$ for each $n < \omega$.
Assume $\pushforward{X^{n \subseteq n+1}} \mu_{n+1} = \mu_n$.
There exists a unique probability measure $\mu_{\omega}$ on $X^{\omega}$ such that $\pushforward{X^{n \subseteq \omega}} \mu_{\omega} = \mu_n$.
\qed
\end{mycorollary}

\section{$K$-th moments of runtimes and rewards}\label{appendix:DetailsandProofsfromsecPreliminaries}

We define a probably measure on the set of runs of a pCFG given a scheduler.
We then define the $k$-th moment of runtimes.
Here we slightly generalize runtime model by considering a reward function and redefine some of the notions to accommodate the reward function.
However, this generalization is not essential, and therefore the readers can safely assume that we are just counting the number of steps until termination (by taking the constant function 1 as a reward function).

Let $\Gamma = (L, V, l_{\init}, \vec{x}_{\init}, {\transition}, \mathrm{Up}, \mathrm{Pr}, G)$ be a pCFG.
A \emph{reward function} on $\Gamma$ is a measurable function $\reward : \configuration \to [0, \infty]$.
Recall that we regard the set $\configuration = L \times \real^V$ of configurations as the product measurable space of $(L, 2^L)$ and $(\real^V, \Borel(\real^V))$.
A \emph{scheduler} of $\Gamma$ resolves two types of nondeterminism: nondeterministic transition and nondeterministic assignment.
\begin{mydefinition}[scheduler]
A \emph{scheduler} $\sigma = (\sigma_t, \sigma_a)$ of $\Gamma$ consists of the following components.
\begin{itemize}
\item A function $\sigma_t : (L \times \real^V)^{*} (L_N \times \real^V) \to \mathcal{D}(L)$ such that
\begin{itemize}
\item if $\pi \in (L \times \real^V)^{*} (L_N \times \real^V)$ and $l \in L_N$ is the last location of $\pi$, then $l' \in \supp(\sigma_t(\pi))$ implies $l \transition l'$, and
\item for each $l \in L$, the mapping $\pi \mapsto \sigma_t(\pi)(\{ l \}) : (L \times \real^V)^{*} (L_N \times \real^V) \to [0, 1]$ is measurable.
\end{itemize}
\item A function $\sigma_a : (L \times \real^V)^{*} (L_{AN} \times \real^V) \to \mathcal{D}(\real)$ such that
\begin{itemize}
\item if $\pi \in (L \times \real^V)^{*} (L_{AN} \times \real^V)$, $l \in L_{AN}$ is the last location of $\pi$ and $(x_j, u) = \mathrm{Up}(l)$, then $\supp(\sigma_a(\pi)) \subseteq u$, and
\item for each $A \in \Borel(\real)$, the mapping $\pi \mapsto \sigma_a(\pi)(A)$ is measurable.
\end{itemize}
\end{itemize}
\end{mydefinition}
Note that if $L_N = \emptyset$ and $L_{AN} = \emptyset$, then there exists only one scheduler that is trivial.

In the rest of the paper, the concatenation of two finite sequences $\rho_1, \rho_2$ is denoted by $\rho_1 \rho_2$ or by $\rho_1 \cdot \rho_2$.

Given a scheduler $\sigma$ and a history of configurations $\rho \in \configuration^{+}$, let $\mu^{\sigma}_{\rho}$ be a probability distribution of the next configurations determined by $\sigma$.
\begin{mydefinition}
Let $\sigma$ be a scheduler and $\rho \in \configuration^{+}$.
A probability measure $\mu^{\sigma}_{\rho}$ on $\configuration$ is defined as follows.
\begin{itemize}
\item If $l \in L_D$ and $\vec{x} \vDash G(l, l')$,
$\mu^{\sigma}_{\rho \cdot (l, \vec{x})} = \delta_{(l', \vec{x})}$.
\item If $l \in L_P$,
$\mu^{\sigma}_{\rho \cdot (l, \vec{x})} = \sum_{l \transition l'} \mathrm{Pr}_{l}(l') \delta_{(l', \vec{x})}$.
\item If $l \in L_N$,
$\mu^{\sigma}_{\rho \cdot (l, \vec{x})} = \sum_{l \transition l'} \sigma_t(\rho \cdot (l, \vec{x}))(\{l'\}) \delta_{(l', \vec{x})}$.
\item Assume $l \in L_A$, $\mathrm{Up}(l) = (x_j, u)$ and $l \transition l'$.
\begin{itemize}
\item If $u \in \Borel(\real^V, \real)$,
$\mu^{\sigma}_{\rho \cdot (l, \vec{x})} = \delta_{(l', \update{\vec{x}}{x_j}{u(\vec{x})})}$.
\item If $u \in \mathcal{D}(\real)$,
$\mu^{\sigma}_{\rho \cdot (l, \vec{x})} = \pushforward{\lambda y. (l', \update{\vec{x}}{x_j}{y})} u$.
\item If $u \in \Borel(\real)$,
$\mu^{\sigma}_{\rho \cdot (l, \vec{x})} = \pushforward{\lambda y. (l', \update{\vec{x}}{x_j}{y})} \sigma_a(\rho \cdot (l, \vec{x}))$.
\end{itemize}
\end{itemize}
\end{mydefinition}

\begin{mylemma}\label{lem:mu_Kleisli}
For each $E \in \Borel(\configuration)$, a mapping $\rho \mapsto \mu^{\sigma}_{\rho}(E) : \configuration^{+} \to [0, 1]$ is measurable.
\end{mylemma}
\begin{proof}
Let $f : \configuration^{+} \to [0, 1]$ be a function defined by $f(\rho) = \mu^{\sigma}_{\rho}(E)$.
It suffices to prove that for each $n < \omega$ and $l \in L$, $f |_{\configuration^n \times (\{ l \} \times \real^V)}: \configuration^n \times (\{ l \} \times \real^V) \to [0, 1]$ is measurable, that is, a function $g_{n, l} : \configuration^n \times \real^V \to [0, 1]$ defined by $g_{n, l}(\rho, \vec{x}) = \mu^{\sigma}_{\rho \cdot (l, \vec{x})}(E)$ is measurable.
\begin{itemize}
\item Assume $l \in L_D$.
\[ g_{n, l}(\rho, \vec{x}) = \delta_{(l, \vec{x})}(E) = 1_E(l, \vec{x}) \]
\item Assume $l \in L_P$.
\[ g_{n, l}(\rho, \vec{x}) = \sum_{l \transition l'} \mathrm{Pr}_l (l') 1_E(l', \vec{x}) \]
\item Assume $l \in L_N$.
\[ g_{n, l}(\rho, \vec{x}) = \sum_{l \transition l'} \sigma_t(\rho \cdot (l, \vec{x}))(\{l'\}) 1_{E}(l', \vec{x}) \]
\item Assume $l \in L_A$, $\mathrm{Up}(l) = (x_j, u)$ and $l \transition l'$.
\begin{itemize}
\item Assume $u \in \Borel(\real^V, \real)$.
\[ g_{n, l}(\rho, \vec{x}) = 1_E (l', \update{\vec{x}}{x_j}{u(\vec{x})}) \]
\item Assume $u \in \mathcal{D}(\real)$.
\[ g_{n, l}(\rho, \vec{x}) = \int_{\real} 1_E (l', \update{\vec{x}}{x_j}{y})\, \mathrm{d} u(y) \]
\item Assume $u \in \Borel(\real)$.
\[ g_{n, l}(\rho, \vec{x}) = \int_{\real} 1_E(l', \update{\vec{x}}{x_j}{y})\, \mathrm{d}(\sigma_a(\rho \cdot (l, \vec{x}))) (y) \]
\end{itemize}
\end{itemize}
In each case, it easily follows that $g_{n, l}$ is measurable.
Note that $\delta_{({-})}(E) = 1_E$ and $(\vec{x}, y) \mapsto \update{\vec{x}}{x_j}{y}$ are measurable functions.
We use Lemma~\ref{lem:Giry_Kleisli} for the last two cases.
\qed
\end{proof}

Given an initial configuration $c_0$, let $\nu^{\sigma}_{c_0, n}$ be a probability measure on the set $\{ c_0 \rho \mid \rho \in \configuration^n \} \cong \configuration^n$ of paths.
\begin{mydefinition}\label{def:measure_finite_path}
For each $n \in \omega$, $\nu^{\sigma}_{c_0, n}$ is a probability measure on $\configuration^n$ defined as
\[ \nu^{\sigma}_{c_0, n}(E) = \begin{cases}
\displaystyle \int_\configuration \dots \int_\configuration 1_{E}(c_1, \dots, c_n)\, \mathrm{d}\mu^{\sigma}_{c_0 \dots c_{n-1}}(c_n) \dots \mathrm{d}\mu_{c_0}(c_1) & \text{if $n > 0$} \\
\delta_{*} & \text{if $n = 0$}
\end{cases} \]
where ${*}$ is the element of $\configuration^0 = \{ {*} \}$.
\end{mydefinition}
Definition~\ref{def:measure_finite_path} is well-defined by Lemma~\ref{lem:Giry_Kleisli} and Lemma~\ref{lem:mu_Kleisli}.

The following lemma is a fundamental property of $\nu^{\sigma}_{c_0, n}$.
\begin{mylemma}
Assume $n > 0$.
For any measurable function $f : \configuration^n \to [0, \infty]$,
\[ \int f\, \mathrm{d}\nu^{\sigma}_{c_0, n} = \int_\configuration \dots \int_\configuration f(c_1, \dots, c_n)\, \mathrm{d}\mu^{\sigma}_{c_0 \dots c_{n-1}}(c_n) \dots \mathrm{d}\mu^{\sigma}_{c_0}(c_1) . \]
\end{mylemma}
\begin{proof}
By the monotone convergence theorem and the linearity of integration.
\qed
\end{proof}

\begin{mylemma}
For each $n \in \mathbb{N}$, $\pushforward{\configuration^{n \subseteq n+1}} \nu^{\sigma}_{c_0, n+1} = \nu^{\sigma}_{c_0, n}$.
\end{mylemma}
\begin{proof}
\begin{align*}
&(\pushforward{\configuration^{n \subseteq n+1}} \nu^{\sigma}_{c_0, n+1}) (E) \\
&= \nu^{\sigma}_{c_0, n+1} (E \times \configuration) \\
&= \int_\configuration \dots \int_\configuration 1_{E \times \configuration}(c_1, \dots, c_{n+1})\, \mathrm{d}\mu^{\sigma}_{c_0 \dots c_{n}}(c_{n+1}) \dots \mathrm{d}\mu^{\sigma}_{c_0}(c_1) \\
&= \int_\configuration \dots \int_\configuration \left( \int_\configuration 1_{\configuration}(c_{n+1})\, \mathrm{d}\mu^{\sigma}_{c_0 \dots c_{n}}(c_{n+1}) \right) \cdot 1_E(c_1, \dots, c_n)\, \mathrm{d}\mu^{\sigma}_{c_0 \dots c_{n-1}}(c_{n}) \dots \mathrm{d}\mu^{\sigma}_{c_0}(c_1) \\
&= \int_\configuration \dots \int_\configuration 1_{E}(c_1, \dots, c_{n})\, \mathrm{d}\mu^{\sigma}_{c_0 \dots c_{n-1}}(c_{n}) \dots \mathrm{d}\mu^{\sigma}_{c_0}(c_1) \\
&= \nu^{\sigma}_{c_0, n} (E)
\tag*{\qed}
\end{align*}
\end{proof}

By Corollary~\ref{cor:Kolmogorov}, we define a probability measure on $\configuration^{\omega}$.
Note that $(\configuration, \Borel(\configuration))$ is a Polish space (a separable completely metrizable topological space), and hence a Radon space.
Therefore, $\nu^{\sigma}_{c_0, n}$ is inner regular.
\begin{mydefinition}
Let $\nu^{\sigma}_{c_0}$ be the probability measure defined as a unique measure such that $\pushforward{\configuration^{n \subseteq \omega}}\nu^{\sigma}_{c_0} = \nu^{\sigma}_{c_0, n}$.
\end{mydefinition}

\begin{mydefinition}[accumulated reward $\reward^{c_0}_C$]
Given a reward function $\reward : \configuration \to [0, \infty]$, let $\reward^{c_0}_C : \configuration^{\omega} \to [0, \infty]$ be a measurable function defined by the sum of the rewards from the initial configuration $c_0$ to the last configuration before entering $C$.
That is,
\[ \reward^{c_0}_C(c_1 c_2 \dots) = \begin{cases}
\sum_{j=0}^{N-1} \reward(c_j) & \text{$\exists N \ge 0$ s.t. $c_N \in C \land (0 \le j < N \implies c_j \notin C)$} \\
\sum_{j=0}^{\infty} \reward(c_j) & \text{otherwise (i.e.\ for each $i$, $c_i \notin C$).}
\end{cases} \]
Note that $\reward(c_0)$ is included in the sum.
\end{mydefinition}

\begin{mydefinition}[$k$-th moment of rewards]
We define two functions $\moment^{\reward, k}_{C, \sigma}, \upper{\moment}^{\reward, k}_{C} : \configuration \to [0, \infty]$ as follows.
\begin{align*}
\moment^{\reward, k}_{C, \sigma}(c_0) &= \int (\reward^{c_0}_C)^k\, \mathrm{d}\nu^{\sigma}_{c_0} & \upper{\moment}^{\reward, k}_{C}(c_0) &= \sup_{\sigma} \moment^{\reward, k}_{C, \sigma}(c_0)
\end{align*}
\end{mydefinition}
Note that $\moment^{\reward, k}_{C, \sigma}$ is measurable by Lemma~\ref{lem:Giry_Kleisli}.

The correspondence of the notations in \S\ref{sec:preliminaries} and in \S\ref{appendix:DetailsandProofsfromsecPreliminaries} is as follows.
\begin{center}
\begin{tabular}{c|c}
\S\ref{sec:preliminaries} & \S\ref{appendix:DetailsandProofsfromsecPreliminaries} \\
\hline
$\nu^{\Gamma}_{\sigma}$ & $\nu^{\sigma}_{c_0}$ where $c_0 = (l_{\init}, \vec{x}_{\init})$ \\
$T^{\Gamma}_{C, \sigma}$ & $\reward^{c_0}_{C}$ where $c_0 = (l_{\init}, \vec{x}_{\init})$ \\
$\moment^{\Gamma,k}_{C, \sigma}$, $\upper{\moment}^{\Gamma,k}_{C}$ & $\moment^{\reward, k}_{C, \sigma}$, $\upper{\moment}^{\reward, k}_{C}$ where $\reward(c) = 1$ for each $c$
\end{tabular}
\end{center}

\section{Omitted Details and Proofs in~\S\ref{sec:supermartingale}}\label{appendix:DetailsandProofsfromsecsupermartingale}
The ultimate goal of this section is to prove Theorem~\ref{thm:uppboundHigher}.
In \S\ref{subsec:preexpectation}-\ref{subsec:timeelapse}, we prove some lemmas regarding to $\upper{\nexttime}$ (Definition~\ref{def:preexpectation}) and $\elapse{1}^{K, k}$ (Definition~\ref{def:timeElapseFuncKth}).
In \S\ref{subsec:highermomentsbyFK}, we prove analogous theorem to Theorem~\ref{thm:uppbound2nd}, Theorem~\ref{thm:second_moment_wo_nondet} and Theorem~\ref{thm:uppboundHigher}.
In \S\ref{subsec:SMHigherMomentsReward}, we prove Theorem~\ref{thm:uppboundHigher}.
We prove them in a generalized way so that an arbitrary reward function is allowed as in \S\ref{appendix:DetailsandProofsfromsecPreliminaries}.

\subsection{Basic properties of the pre-expectation}\label{subsec:preexpectation}
We prove several lemmas for $\upper{\nexttime}$ in Definition~\ref{def:preexpectation}.

The next lemma claims that we can ignore outside of an invariant $I$.
\begin{mylemma}\label{lem:nexttime_invariant}
Let $I$ be an invariant.
If $\eta(c) = \eta'(c)$ for any $c \in I$, then $(\upper{\nexttime} \eta)(c) = (\upper{\nexttime} \eta')(c)$ for any $c \in I$.
\qed
\end{mylemma}




The complete lattice $[0, \infty]$ has the following properties as an $\omega$-cpo, and the set of functions $\configuration \to [0, \infty]^K$ inherits the same properties.
\begin{itemize}
\item Let $\{ \eta_n \}_{n < \omega}$ and $\{ \eta'_n \}_{n < \omega}$ be $\omega$-chains.
Then we have
\begin{equation}
\sup_{n \in \omega} \eta_n + \sup_{n \in \omega} \eta'_n = \sup_{n \in \omega} (\eta_n + \eta'_n).
\label{eq:sup_add}
\end{equation}
That is, the addition ${+}$ is $\omega$-continuous.
\item Let $\{ \eta_n \}_{n < \omega}$ be a $\omega$-chain and $a \ge 0$.
Then we have
\begin{equation}
a \cdot \sup_{n \in \omega} \eta_n = \sup_{n \in \omega} (a \cdot \eta_n).
\label{eq:sup_scalar_mult}
\end{equation}
That is, $a \cdot ({-})$ is $\omega$-continuous.
\end{itemize}
These properties are often used in the proofs of $\omega$-continuity in the rest of the paper.

\begin{mylemma}\label{lem:nexttime_cont}
$\upper{\nexttime}$ is $\omega$-continuous.
\end{mylemma}
\begin{proof}
Let $\{ \eta_n : \configuration \to [0, \infty] \}_{n \in \omega}$ be an $\omega$-chain.
We prove $(\upper{\nexttime} (\sup_{n \in \omega} \eta_n)) (l, \vec{x}) = \sup_{n \in \omega} (\upper{\nexttime}) (l, \vec{x})$ for each $(l, \vec{x}) \in L \times \real^V$.
\begin{itemize}
\item Assume $l \in L_D$ and $\vec{x} \vDash G(l, l')$.
\[ (\upper{\nexttime} (\sup_{n \in \omega} \eta_n)) (l, \vec{x}) = (\sup_{n \in \omega} \eta_n)(l', \vec{x}) = \sup_{n \in \omega} (\eta_n(l', \vec{x})) = \sup_{n \in \omega} (\upper{\nexttime} \eta_n) (l, \vec{x}) \]
\item Assume $l \in L_P$.
\begin{align*}
(\upper{\nexttime} (\sup_{n \in \omega} \eta_n)) (l, \vec{x}) &= \sum_{l \transition l'} \mathrm{Pr}_l(l') (\sup_{n \in \omega} \eta_n)(l', \vec{x}) \\
&= \sup_{n \in \omega} \sum_{l \transition l'} \mathrm{Pr}_l(l') \eta_n(l', \vec{x}) \\
&= \sup_{n \in \omega} (\upper{\nexttime} \eta_n) (l, \vec{x})
\end{align*}
\item Assume $l \in L_N$.
\[ (\upper{\nexttime} (\sup_{n \in \omega} \eta_n)) (l, \vec{x}) = \sup_{l \transition l'} \sup_{n \in \omega} \eta_n(l', \vec{x}) = \sup_{n \in \omega} \sup_{l \transition l'} \eta_n(l', \vec{x}) = \sup_{n \in \omega} (\upper{\nexttime} \eta_n) (l, \vec{x}). \]
\item Assume $l \in L_A$, $\mathrm{Up}(l) = (x_j, u)$ and $l \transition l'$.
\begin{itemize}
\item Assume $u \in \Borel(\real^V, \real)$.
\[ (\upper{\nexttime} (\sup_{n \in \omega} \eta_n)) (l, \vec{x}) = \sup_{n \in \omega} \eta_n(l', \update{\vec{x}}{x_j}{u(\vec{x})}) = \sup_{n \in \omega} (\upper{\nexttime} \eta_n) (l, \vec{x}) \]
\item Assume $u \in \mathcal{D}(\real)$.
\begin{align*}
(\upper{\nexttime} (\sup_{n \in \omega} \eta_n)) (l, \vec{x}) &= \int_{\real} (\sup_{n \in \omega} \eta_n)(l', \update{\vec{x}}{x_j}{y})\, \mathrm{d} u(y) \\
&= \sup_{n \in \omega} \int_{\real} \eta_n(l', \update{\vec{x}}{x_j}{y})\, \mathrm{d} u(y) \\
&= \sup_{n \in \omega}  (\upper{\nexttime} \eta_n) (l, \vec{x})
\end{align*}
by the monotone convergence theorem.
\item Assume $u \in \Borel(\real)$. 
\begin{align*}
(\upper{\nexttime} (\sup_{n \in \omega} \eta_n)) (l, \vec{x}) &= \sup_{y \in u} \sup_{n \in \omega} \eta_n(l', \update{\vec{x}}{x_j}{y}) \\
&= \sup_{n \in \omega} \sup_{y \in u} \eta_n(l', \update{\vec{x}}{x_j}{y}) \\
&= \sup_{n \in \omega} (\upper{\nexttime} \eta_n) (l, \vec{x})
\tag*{\qed}
\end{align*}
\end{itemize}
\end{itemize}
\end{proof}

The next lemma is a justification of the name ``pre-expectation''.
\begin{mylemma}\label{lem:nexttime}
For any configuration $c_0$, measurable function $\eta : \configuration \to [0, \infty]$, scheduler $\sigma$,
\[ \upper{\nexttime}\eta (c_0) \ge \int_\configuration \eta(c_1) d\mu^{\sigma}_{c_0}(c_1) . \]
\end{mylemma}
\begin{proof}
Let $(l_0, \vec{x}_0) = c_0$.
\begin{itemize}
\item Assume $l_0 \in L_D$ and $\vec{x}_0 \vDash G(l_0, l_1)$.

\[ \int_\configuration \eta(c_1) d\mu^{\sigma}_{c_0}(c_1) = \eta(l_1, \vec{x}_0) = \upper{\nexttime}\eta (c_0) \]

\item Assume $l_0 \in L_P$.

\[ \int_\configuration \eta(c_1) d\mu^{\sigma}_{c_0}(c_1) = \sum_{l_0 \transition l_1} \mathrm{Pr}_{l_0}(l_1) \eta(l_1, \vec{x}_0) = \upper{\nexttime}\eta (c_0) \]

\item Assume $l_0 \in L_N$.
\[ \int_\configuration \eta(c_1) d\mu^{\sigma}_{c_0}(c_1) = \sum_{l_0 \transition l_1} \sigma_t(c_0)(l_1) \eta(l_1, \vec{x}_0) \le \sup_{l_0 \transition l_1} \eta(l_1, \vec{x}_0) = \upper{\nexttime}\eta (c_0) \]

\item Assume $l_0 \in L_A$, $\mathrm{Up}(l_0) = (x_j, u)$ and $l_0 \transition l_1$.
\begin{itemize}

\item Assume $u \in \Borel(\real^V, \real)$.
\[ \int_\configuration \eta(c_1) d\mu^{\sigma}_{c_0}(c_1) = \eta(l_1, \update{\vec{x}_0}{x_j}{u(\vec{x}_0)}) = \upper{\nexttime}\eta (c_0) \]

\item Assume $u \in \mathcal{D}(\real)$.
\[ \int_\configuration \eta(c_1) d\mu^{\sigma}_{c_0}(c_1) = \int_{\real} \eta(l_1, \update{\vec{x}_0}{x_j}{y}) du(y) = \upper{\nexttime}\eta (c_0) \]

\item Assume $u \in \Borel(\real)$.
\begin{align*}
\int_\configuration \eta(c_1) d\mu^{\sigma}_{c_0}(c_1) &= \int_{\real} \eta(l_1, \update{\vec{x}_0}{x_j}{y}) d(\sigma_a(c_0))(y) \\
&\le \sup_{y \in u} \eta(l_1, \update{\vec{x}_0}{x_j}{y}) = \upper{\nexttime}\eta (c_0)
\end{align*}
\end{itemize}
\end{itemize}
\qed
\end{proof}

\begin{mylemma}\label{lem:nexttime_wo_nondet}
Assume $L_N = \emptyset$ and $L_{AN} = \emptyset$ and let $\sigma$ be the unique scheduler that plays no role.
For any configuration $c_0$ and any measurable function $\eta : \configuration \to [0, \infty]$,
\[ \upper{\nexttime}\eta (c_0) = \int_\configuration \eta(c_1) d\mu^{\sigma}_{c_0}(c_1). \]
\end{mylemma}
\begin{proof}
Immediate from the proof of Lemma~\ref{lem:nexttime}.
\qed
\end{proof}

\subsection{Basic properties of the time-elapse function}\label{subsec:timeelapse}
We next prove lemmas for the time-elapse function in Definition~\ref{def:timeElapseFuncKth}.
We redefine the time-elapse function for the generalized runtime model.
\begin{mydefinition}[time-elapse function]
For each $a \in [0, \infty]$, natural number $K$ and $k \in \{ 1, \dots, K \}$, $\elapse{a}^{K, k} : [0, \infty]^K \to [0, \infty]$ is a function defined by
\[ \elapse{a}^{K, k}(x_1, \dots, x_K) = a^k + \sum_{j=1}^k \binom{k}{j} a^{k-j} x_j \]
\end{mydefinition}

\begin{mylemma}\label{lem:elapse_cont}
$\elapse{a}^{K, k}$ is $\omega$-continuous.
\end{mylemma}
\begin{proof}
Immediate from (\ref{eq:sup_add}) and (\ref{eq:sup_scalar_mult}).
\qed
\end{proof}

\begin{mylemma}[commutativity of $\int$ and $\elapse{a}^{K,k}$]
For any probability measure $\mu$ on $X$, any measurable functions $f_1, \dots, f_n : X \to [0, \infty]$ and $a \in [0, \infty]$, $\elapse{a}^{k, n}$ and integrals commute.
That is
\[ \int \elapse{a}^{K, k}(f_1(x), \dots, f_n(x)) d\mu(x) = \elapse{a}^{K, k} \left( \int f_1(x) d\mu(x), \dots, \int f_n(x) d\mu(x) \right). \]
\end{mylemma}
\begin{proof}
By the linearity of integration.
\qed
\end{proof}

\subsection{Characterizing higher moments by $F_K$}\label{subsec:highermomentsbyFK}
We prove Theorem~\ref{thm:uppbound2nd}, Theorem~\ref{thm:second_moment_wo_nondet} and Theorem~\ref{thm:uppboundHigher} in the generalized runtime model.
We first extend Definition~\ref{def:FK} so that an arbitrary reward is allowed.
\begin{mydefinition}
Let $I$ be an invariant and $C \subseteq I$ be a Borel set.
Let $F_K : (\configuration \to [0, \infty]^K) \to (\configuration \to [0, \infty]^K)$ be a function defined by $F_K(c) = (F_{K, 1}(c), \dots, F_{K, K}(c))$ where the $k$-th component $F_{K, k} : (\configuration \to [0, \infty]^K) \to (\configuration \to [0, \infty])$ of $F_K$ is defined by
\[ F_{K, k} (\eta) (c) = \begin{cases}
(\upper{\nexttime}(\elapse{\reward(c)}^{K, k} \comp \eta))(c) & c \in I \setminus C \\
0 & \text{otherwise.} \\
\end{cases} \]
\end{mydefinition}

\begin{mylemma}\label{lem:FK_cont}
$F_K$ is $\omega$-continuous.
\end{mylemma}
\begin{proof}
Immediate from Lemma~\ref{lem:nexttime_cont} and Lemma~\ref{lem:elapse_cont}.
\qed
\end{proof}

The following theorems generalizes Theorem~\ref{thm:uppbound2nd},\ref{thm:uppboundHigher} and Theorem~\ref{thm:second_moment_wo_nondet}, respectively.
\begin{mytheorem}\label{thm:uppboundHigherReward}
\[ \lfp F_K \ge \left\langle \upper{\moment}^{\reward, 1}_{C}, \dots, \upper{\moment}^{\reward, K}_{C} \right\rangle  \]
for any $c_0 \in I$.
\end{mytheorem}

\begin{mytheorem}\label{thm:uppboundHigherRewardNondet}
If $L_N = \emptyset$ and $L_{AN} = \emptyset$,
\[ \lfp F_K = \left\langle \upper{\moment}^{\reward, 1}_{C}, \dots, \upper{\moment}^{\reward, K}_{C} \right\rangle  \]
for any $c_0 \in I$.
\end{mytheorem}

Here a function $\langle f_1, \dots, f_n \rangle$ is defined by $\langle f_1, \dots, f_K \rangle (x) = (f_1(x), \dots, f_K(x))$.

To prove Theorem~\ref{thm:uppboundHigherReward} and Theorem~\ref{thm:uppboundHigherRewardNondet}, we consider an approximation of $k$-th moments of accumulated rewards up to finite steps.

\begin{mydefinition}[accumulated reward up to $n$ steps]
Let $\reward^{c_0}_{C, n} : \configuration^n \to [0, \infty]$ be a measurable function defined by
\[ \reward^{c_0}_{C, n}(c_1 \dots c_n) = \begin{cases}
\sum_{j=0}^{N-1} \reward(c_j) & \text{$\exists N \ge 0$. $c_i \in C \land (0 \le j < N \implies c_j \notin C)$} \\
\sum_{j=0}^{n-1} \reward(c_j) & \text{otherwise}
\end{cases} \]
\end{mydefinition}
The definition of $\reward^{c_0}_{C, n}$ is similar to $\reward^{c_0}_{C}$ except that the sum of the value of reward function is restricted to the first $n$ configurations.
The next lemma shows a connection between $\reward^{c_0}_{C}$ and $\reward^{c_0}_{C, n}$.
\begin{mylemma}\label{lem:rew_limit}
$\{ \reward^{c_0}_{C, n} \comp \configuration^{n \subseteq \omega} : \configuration^{\omega} \to [0, \infty] \}_n$ is an increasing sequence of functions and its limit is $\reward^{c_0}_{C}$.
\end{mylemma}
\begin{proof}
Given $\rho = c_1 c_2 \dots \in \configuration^{\omega}$, there are two cases.
\begin{itemize}
\item Assume there exists $N \in \omega$ such that $c_N \in C$ and $0 \le j < N \implies c_j \notin C$.
\begin{align*}
\reward^{c_0}_{C, n} \comp \configuration^{n \subseteq \omega}(c_1 c_2 \dots) &= \begin{cases}
\displaystyle \sum_{j=0}^{n-1} \reward(c_j) & \text{if } n < N - 1 \\
\displaystyle \sum_{j=0}^{N-1} \reward(c_j) & \text{if } N - 1 \le n
\end{cases} \\
\reward^{c_0}_{C}(c_1 c_2 \dots) &= \sum_{j=0}^{N-1} \reward(c_j)
\end{align*}
\item Assume $\rho \in (\configuration \setminus C)^{\omega}$.
\begin{align*}
\reward^{c_0}_{C, n} \comp \configuration^{n \subseteq \omega}(c_1 c_2 \dots) &= \sum_{j=0}^{n-1} \reward(c_j) \\
\reward^{c_0}_{C} (c_1 c_2 \dots) &= \sum_{j=0}^{\infty} \reward(c_j)
\end{align*}
\end{itemize}
In both cases, it is easy to prove
$\reward^{c_0}_{C, n} \comp \configuration^{n \subseteq \omega} \le \reward^{c_0}_{C, n+1} \comp \configuration^{n+1 \subseteq \omega}$
for each $n$, and
$\reward^{c_0}_{C} = \sup_{n \in \omega} \big( \reward^{c_0}_{C, n} \comp \configuration^{n \subseteq \omega} \big)$.
\qed
\end{proof}

The $k$-th moment of $\reward^{c_0}_{C, n}$ is denoted by $\moment^{\reward, k}_{C, \sigma, n}(c_0)$.
\begin{mydefinition}[$k$-th moment up to $n$ steps]
A function $\moment^{\reward, k}_{C, \sigma, n} : \configuration \to [0, \infty]$ is defined as follows.
\[ \moment^{\reward, k}_{C, \sigma, n}(c_0) = \int (\reward^{c_0}_{C, n})^k d\nu^{\sigma}_{c_0, n} \]
\end{mydefinition}

A connection between $\moment^{\reward, k}_{C, \sigma, n}$ and $\moment^{\reward, k}_{C, \sigma}$ is as follows.
\begin{mylemma}
A sequence $\{ \moment^{\reward, k}_{C, \sigma, n} \}_{n \in \omega}$ is increasing and its limit is $\moment^{\reward, k}_{C, \sigma}$:
\[ \moment^{\reward, k}_{C, \sigma} = \sup_{n \in \omega} \moment^{\reward, k}_{C, \sigma, n}. \]
\end{mylemma}
\begin{proof}
The former part is immediate by Lemma~\ref{lem:rew_limit}.
The latter part is proved by the following calculation.
\begingroup
\allowdisplaybreaks
\begin{align*}
&\moment^{\reward, k}_{C, \sigma}(c_0) \\
&= \int (\reward^{c_0}_{C})^k\, \mathrm{d}\nu^{\sigma}_{c_0} \\
&= \int \sup_{n \in \omega} (\reward^{c_0}_{C, n})^k \comp \configuration^{n \subseteq \omega}\, \mathrm{d}\nu^{\sigma}_{c_0} & \text{(by Lemma~\ref{lem:rew_limit})}\\
&= \sup_{n \in \omega} \int (\reward^{c_0}_{C, n})^k \comp \configuration^{n \subseteq \omega}\, \mathrm{d}\nu^{\sigma}_{c_0} & \text{(by the monotone convergence theorem)} \\
&= \sup_{n \in \omega} \int (\reward^{c_0}_{C, n})^k\, \mathrm{d}\big(\pushforward{\configuration^{n \subseteq \omega}} \nu^{\sigma}_{c_0} \big) & \text{(by Lemma~\ref{lem:pushforward})} \\
&= \sup_{n \in \omega} \int (\reward^{c_0}_{C, n})^k\, \mathrm{d}\nu^{\sigma}_{c_0, n} \\
&= \sup_{n \in \omega} \moment^{\reward, k}_{C, \sigma, n}(c_0) \tag*{\qed}
\end{align*}
\endgroup
\end{proof}

\begin{mydefinition}
For any $c$ and $\sigma$, we define a scheduler $\sigma^{c} = (\sigma^{c}_t, \sigma^{c}_a)$ by $\sigma^{c}_t (\rho) = \sigma_t(c \rho)$ and $\sigma^{c}_a (\rho) = \sigma_a(c \rho)$.
\end{mydefinition}

The following lemma easily follows from the definition of $\mu^{\sigma}_{\rho}$.
\begin{mylemma}
$\mu^{\sigma^{c_0}}_{\rho} = \mu^{\sigma}_{c_0 \rho}$
\qed
\end{mylemma}


The following lemma expresses the $n+1$ step approximation $\moment^{\reward, k}_{C, \sigma, n+1}$ in terms of the $n$ step approximations $\moment^{\reward, 1}_{C, \sigma^{c_0}, n}, \dots, \moment^{\reward, K}_{C, \sigma^{c_0}, n}$, which plays a crucial role in the induction step in the proof of Theorem~\ref{thm:uppboundHigherReward} and Theorem~\ref{thm:uppboundHigherRewardNondet}.
\begin{mylemma}
\label{lem:moment_induction_kth}
Assume $c_0 \notin C$ and $k \in \{ 1, \dots, K \}$.
For each $n \in \omega$,
\[ \moment^{\reward, k}_{C, \sigma, n+1}(c_0) = \elapse{\reward(c_0)}^{K, k} \left( \int_\configuration \moment^{\reward, 1}_{C, \sigma^{c_0}, n}(c_1)\, \mathrm{d}\mu^{\sigma}_{c_0}(c_1), \dots, \int_\configuration \moment^{\reward, K}_{C, \sigma^{c_0}, n}(c_1)\, \mathrm{d}\mu^{\sigma}_{c_0}(c_1) \right). \]
\end{mylemma}
\begin{proof}
\begingroup
\allowdisplaybreaks
\begin{align*}
&\moment^{\reward, k}_{C, \sigma, n+1}(c_0) \\
&= \int_{\configuration^{n+1}} (\reward^{c_0}_{C, n+1})^k\, \mathrm{d}\nu^{\sigma}_{c_0, n+1} \\
&= \int_\configuration \dots \int_\configuration \left( \reward^{c_0}_{C, n+1}(c_1, \dots, c_{n+1}) \right)^k\, \mathrm{d}\mu^{\sigma}_{c_0 \dots c_{n}}(c_{n+1}) \dots \mathrm{d}\mu^{\sigma}_{c_0}(c_1) \\
&= \int_\configuration \dots \int_\configuration \left( \reward(c_0) + \reward^{c_1}_{C, n}(c_2, \dots, c_{n+1}) \right)^k\, \mathrm{d}\mu^{\sigma}_{c_0 \dots c_{n}}(c_{n+1}) \dots \mathrm{d}\mu^{\sigma}_{c_0}(c_1) \\
&= \int_\configuration \dots \int_\configuration \left( \sum_{j=0}^k \binom{k}{j} (\reward(c_0))^{k-j} \left(\reward^{c_1}_{C, n}(c_2, \dots, c_{n+1}) \right)^j \right)\, \mathrm{d}\mu^{\sigma}_{c_0 \dots c_{n}}(c_{n+1}) \dots \mathrm{d}\mu^{\sigma}_{c_0}(c_1) \\
&= (\reward(c_0))^k + \sum_{j=1}^k \binom{k}{j} (\reward(c_0))^{k-j} \\
&\qquad \cdot \int_\configuration \left( \int_\configuration \dots \int_\configuration \left( \reward^{c_1}_{C, n}(c_2, \dots, c_{n+1}) \right)^j\, \mathrm{d}\mu^{\sigma^{c_0}}_{c_1 \dots c_{n}}(c_{n+1}) \dots \mathrm{d}\mu^{\sigma^{c_0}}_{c_1}(c_2) \right) \mathrm{d}\mu^{\sigma}_{c_0}(c_1) \\
&= (\reward(c_0))^k + \sum_{j=1}^k \binom{k}{j} (\reward(c_0))^{k-j} \int_\configuration \int_{\configuration^n} (\reward^{c_1}_{C, n})^j\, \mathrm{d}\nu^{\sigma^{c_0}}_{c_1, n} \mathrm{d}\mu^{\sigma}_{c_0}(c_1) \\
&= (\reward(c_0))^k + \sum_{j=1}^k \binom{k}{j} (\reward(c_0))^{k-j} \int_\configuration \moment^{\reward, j}_{C, \sigma^{c_0}, n}(c_1)\, \mathrm{d}\mu^{\sigma}_{c_0}(c_1)
\tag*{\qed}
\end{align*}
\endgroup
\end{proof}

\begin{proof}[Theorem~\ref{thm:uppboundHigherReward}]\mbox{}
\begin{enumerate}
\item
We prove
\begin{equation}
(F_K)^n(\bot) \ge \left\langle \moment^{\reward, 1}_{C, \sigma, n}, \dots, \moment^{\reward, K}_{C, \sigma, n} \right\rangle
\label{eq:FKn_ge_moments_up_to_n}
\end{equation}
for each $\sigma$ and $n$ by induction on $n$.

\begin{itemize}
\item If $n = 0$, the l.h.s.\ and the r.h.s.\ are equal to 0.
\item If $n > 0$, it suffices to prove that for each $c_0$, there exists $\sigma'$ such that
\[ F_K \left( \left\langle \moment^{\reward, 1}_{C, \sigma', n}, \dots, \moment^{\reward, K}_{C, \sigma', n} \right\rangle \right) (c_0) \ge \left(\moment^{\reward, 1}_{C, \sigma, n+1}(c_0), \dots, \moment^{\reward, K}_{C, \sigma, n+1}(c_0) \right) \]
by the induction hypothesis.
If $c_0 \in C$, the l.h.s.\ and the r.h.s.\ are equal to 0.
If $c_0 \notin C$, we prove
\[ \upper{\nexttime} \left(\elapse{\reward(c_0)}^{K, k} \comp \left\langle \moment^{\reward, 1}_{C, \sigma^{c_0}, n}, \dots, \moment^{\reward, K}_{C, \sigma^{c_0}, n} \right\rangle \right) (c_0) \ge \moment^{\reward, k}_{C, \sigma, n+1}(c_0). \]
By Lemma~\ref{lem:moment_induction_kth}, it suffices to prove
\[ \upper{\nexttime}\eta (c_0) \ge \int_\configuration \eta(c_1)\, \mathrm{d}\mu^{\sigma}_{c_0}(c_1) \]
where
\[ \eta = \elapse{\reward(c_0)}^{K, k} \comp \left( \moment^{\reward, 1}_{C, \sigma^{c_0}, n}, \dots, \moment^{\reward, K}_{C, \sigma^{c_0}, n} \right) . \]
This holds by Lemma~\ref{lem:nexttime}.
\end{itemize}
\item We take supremum of (\ref{eq:FKn_ge_moments_up_to_n}) with respect to $n$, and then with respect to $\sigma$.
\begin{equation}
\lfp F_K \ge \sup_{n \in \omega} \big( (F_K)^n(\bot) \big) \ge \left\langle \upper{\moment}^{\reward, 1}_{C}, \dots, \upper{\moment}^{\reward, K}_{C} \right\rangle
\tag*{\qed}
\end{equation}
\end{enumerate}
\end{proof}

\begin{proof}[Theorem~\ref{thm:uppboundHigherRewardNondet}]
Here, we prove
\[ (F_K)^n(\bot) = \left\langle \moment^{\reward, 1}_{C, \sigma, n}, \dots, \moment^{\reward, K}_{C, \sigma, n} \right\rangle \]
for each $n$ by induction on $n$ in the same way as Theorem~\ref{thm:uppboundHigherReward} except that we use Lemma~\ref{lem:nexttime_wo_nondet} instead of Lemma~\ref{lem:nexttime}.

By the Kleene fixed-point theorem and Lemma~\ref{lem:FK_cont}, we have $\sup_{n \in \omega} \big( (F_K)^n (\bot) \big) = \lfp F_K$.

\begin{align*}
\lfp F_K &= \sup_{n \in \omega} \big( (F_K)^n (\bot) \big) \\
&= \sup_{n \in \omega} \left\langle \moment^{\reward, 1}_{C, \sigma, n}, \dots, \moment^{\reward, K}_{C, \sigma, n} \right\rangle \\
&= \left\langle \moment^{\reward, 1}_{C, \sigma}, \dots, \moment^{\reward, K}_{C, \sigma} \right\rangle
\end{align*}
Since there is only one scheduler if $L_N = L_{AN} = \emptyset$, we conclude
\begin{equation}
\mu F_K = \left\langle \moment^{\reward, 1}_{C, \sigma}, \dots, \moment^{\reward, K}_{C, \sigma} \right\rangle = \left\langle \upper{\moment}^{\reward, 1}_{C}, \dots, \upper{\moment}^{\reward, K}_{C} \right\rangle .
\tag*{\qed}
\end{equation}
\end{proof}

\subsection{Ranking supermartingale for $K$-th moments}\label{subsec:SMHigherMomentsReward}
The following definition and theorem generalize Definition~\ref{def:ranksupHigher} and Theorem~\ref{thm:uppboundHigher}, respectively.
\begin{mydefinition}[ranking supermartingale for $K$-th moments of accumulated rewards]\label{def:generalranksupHigher}
A ranking supermartingale for $K$-th moments is a function $\eta : \configuration \to \real^K$ such that for each $k$,
\begin{itemize}
\item $\eta_k(c) \geq (\upper{\nexttime} (\elapse{\reward(c)}^{K, k} \comp \eta_k))(c)$ for each $c \in I \setminus C$
\item $\eta_k(c) \geq 0$ for each $c \in I$
\end{itemize}
where $\eta_k : \configuration \to \real$ is defined by $(\eta_1(c), \dots, \eta_K(c)) = \eta(c)$ for each $c \in \configuration$.
\end{mydefinition}

\begin{mytheorem}\label{thm:uppboundHigherSMReward}
If $\eta$ is a supermartingale for $K$-th moments, then for any $c \in I$, $(\upper{\moment}^{\reward, 1}_{C}(c), \dots, \upper{\moment}^{\reward, K}_{C}(c)) \leq \eta(c)$.
\end{mytheorem}
\begin{proof}
Let $\eta' : \configuration \to [0, \infty]^K$ be a function defined by
\[ \eta'(c) = \begin{cases}
\eta(c) & c \in I \\
0 & \text{otherwise.}
\end{cases} \]
By Lemma~\ref{lem:nexttime_invariant}, $F_K(\eta') \le \eta'$ is easily proved.
By the Knaster-Tarski theorem, we have $\lfp F_K \le \eta'$.
Therefore
\[ (\upper{\moment}^{\reward, 1}_{C}(c), \dots, \upper{\moment}^{\reward, K}_{C}(c)) \le \lfp F_K (c) \le \eta'(c) = \eta(c) \]
for each $c \in I$.
\qed
\end{proof}

\section{Details of Template-Based Synthesis Algorithm}\label{sec:synthesisApp}
In this section we describe the template-based synthesis algorithms in \S{}\ref{sec:synthesis} in more detail.
They are based on existing template-based algorithms for synthesizing a ranking supermartingale for first moments 
in~\cite{DBLP:conf/cav/ChakarovS13,DBLP:journals/toplas/ChatterjeeFNH18,DBLP:conf/cav/ChatterjeeFG16}.
Input to the algorithm is a pCFG $\Gamma$,  an invariant $I$, a set $C \subseteq I$ of configurations, and a natural number $K$.
Output is an upper bound of $K$-th moment.

\subsection{Linear Template-based Algorithm}
Synthesis of a ranking supermartingale via reduction to an LP problem is discussed in~\cite{DBLP:conf/cav/ChakarovS13,DBLP:journals/toplas/ChatterjeeFNH18}.
We adapt this for our supermartingales.

We first define some notions.

\begin{mydefinition}\label{def:linexp}
Let $V=\{x_1,\ldots,x_n\}$ be a set of variables. 
A \emph{linear expression} over $V$ is a formula of a form $a_1x_{i_1}+\cdots+a_nx_{i_n}+b$ where $a_1,\ldots,a_n,b\in\mathbb{R}$ and $x_{i_1},\ldots,x_{i_n}\in V$.
We write $\linexpr{\real}{V}$ for the set of linear expressions. 
A \emph{linear inequality} over $V$ is a formula of a form $\varphi\geq 0$ where $\varphi$ is a linear expression.
A \emph{linear conjunctive predicate} is a conjunction $\varphi_1\geq 0\wedge \cdots \wedge \varphi_p\geq 0$ of linear constraints,
and a \emph{linear predicate} is a disjunction 
$(\varphi_{1,1}\geq 0\wedge \cdots \wedge \varphi_{1,p_1}\geq 0)\vee\cdots\vee(\varphi_{q,1}\geq 0\wedge \cdots \wedge \varphi_{q,p_q}\geq 0)$
of linear conjunctive predicates. 
We define their semantics in the standard manner. 

For a pCFG $\Gamma=(L, V, l_{\init}, \vec{x}_{\init}, {\transition}, \mathrm{Up}, \mathrm{Pr}, G)$,
a \emph{linear expression map} (resp.\ \emph{linear predicate map}) over $\Gamma$ is a function that assigns 
a linear expression (resp.\ linear predicate) to each location of $\Gamma$. 
The semantics of the former is a function assigning a real number to each configuration, i.e.\ it has a type $L\times \mathbb{R}^V\to \mathbb{R}$, and 
that of the latter is a set of configurations, i.e.\ a subset of $L\times\mathbb{R}^V$.
They are defined in the natural manners.
\end{mydefinition}

In the rest of this section, we describe a linear template-based synthesis algorithm for a pCFG $\Gamma$
an invariant $I$, a set $C \subseteq I$ of configurations, and a natural number $K$.
We assume that the input satisfies the following conditions.
Similar conditions were assumed in~\cite{DBLP:conf/cav/ChakarovS13,DBLP:journals/toplas/ChatterjeeFNH18}.

\begin{myassumption}\mbox{}
\begin{itemize}
\item For any $l \in L_A$ such that $\mathrm{Up}(l) = (x_j, u)$,
\begin{itemize}
\item if $u \in \Borel(\real^V, \real)$, then $u$ is represented by a linear expression over $V$;
\item if $u \in \mathcal{D}(\real)$, the expectation of $u$ is known; and
\item if $u \in \Borel(\real)$, then $u$ is represented by a linear predicate $\phi$ over $\{ x_j \}$.
\end{itemize}
\item For any $l \in L_D$ and $l' \in L$, $G(l, l') = \llbracket p \rrbracket$ is represented by a linear predicate over $V$.
\item the invariant $I$ is represented by a linear predicate map over $\Gamma$.
\item the set $C$ of terminal configurations is represented by a linear conjunctive predicate map.
\end{itemize}
\end{myassumption}

Let $V=\{x_1,\ldots,x_n\}$ be the set of variables appearing in $\Gamma$.
We first fix a \emph{linear template} to a supermartingale.
It is a family of formulas indexed by $i\in\{1,\ldots,K\}$ and $l\in L$ that have the following form:
\[
\eta_i(l, \vec{x}) \;=\;
a^l_{1,i}x_1+\cdots +a^l_{n,i}x_n + b^l_{i}\,.
\]
Here $a^l_{1,i},\ldots,a^l_{n,i},b^l_{i}$ are newly added variables called \emph{parameters}.
We write $U$ for the set of all parameters, i.e.\ $U:=\{a^l_{1,i},\ldots,a^l_{n,i},b^l_{i}\mid i\in\{1,\ldots,K\},l\in L\}$.
Note that if we fix a valuation $U\to\mathbb{R}$ of parameters, 
then each $\eta_i(l, \vec{x})$ reduces to a linear expression over $V$,
and therefore $\eta_i(\place, \vec{x})$ can be regarded as a linear expression map $L\times \real^V\to \linexpr{\real}{V}$.
Our goal is to find a valuation $U\to\mathbb{R}$ so that a $K$-tuple 
$\bigl(\eta_1(\place, \vec{x}),\ldots,\eta_K(\place, \vec{x})\bigr)$
of linear expression maps become 
a ranking supermartingale for $K$-th moment (Definition~\ref{def:ranksupHigher}).

%

To this end, we reduce the axioms of ranking supermartingale for $K$-th moments in Definition~\ref{def:ranksupHigher} to 
conditions over the parameters.
We shall omit the detail, but it is not so hard to see that 
as a result of the reduction we obtain a conjunction of formulas of the following form:
\begin{equation}\label{eq:1811101315}
\forall \vec{x}\in\real^V.\; \varphi_1\rhd_1 0\wedge\cdots \varphi_m\rhd_m 0 \implies \psi\geq 0\,.
\end{equation}
Here $\rhd_i\in \{\geq,>\}$, each $\varphi_i$ is a linear expression without parameters, and 
$\psi$ is a linear formula over $V$ whose coefficients are linear expressions over $U$.

We next relax the strict inequalities as follows: 
\begin{equation}\label{eq:1811101316}
\forall \vec{x}\in\real^V.\; \varphi_1\geq 0\wedge\cdots \varphi_m\geq 0 \implies \psi\geq 0\,.
\end{equation}
It is easy to see that (\ref{eq:1811101316}) implies (\ref{eq:1811101315}).
The same relaxation is also done in~\cite{DBLP:conf/cav/ChakarovS13,DBLP:journals/toplas/ChatterjeeFNH18}.

Using matrices, we can represent a formula~(\ref{eq:1811101316}) as follows:
\begin{equation}\label{eq:1811101317}
\forall \vec{x}\in\real^V.\; 
A \vec{x} \leq b \implies \transpose{c} x \leq d\,.
\end{equation}
Here $A$ is a matrix and $b$ is a vector all of whose components are real numbers,
and $c$ is a vector and $d$ is a scalar all of whose components are linear expressions over $U$.
In~\cite{DBLP:conf/cav/ChakarovS13,DBLP:journals/toplas/ChatterjeeFNH18}, a formula (\ref{eq:1811101317}) is reduced to the following formula:
\begin{equation}\label{eq:1811101318}
\exists \vec{y}\in\real^m.\; 
\exists y'\in\real.\;
d-\transpose{c} x = \transpose{\vec{y}}\bigl(b-A\vec{x}\bigr)+y'\,.
\end{equation}
Here $m$ is the dimension of $b$.
It is easy to see that (\ref{eq:1811101318}) implies (\ref{eq:1811101317}).
By comparing the coefficients on both sides, we can see that (\ref{eq:1811101318}) is equivalent to 
\[
\exists \vec{y} \in \real^m.\; \transpose{A} \vec{y} = c \land \transpose{b} y \le d\,.
\]
Note that the resulting (in)equalities are linear with respect to parameters in $U$ and $\vec{y}$. 
Hence its feasibility can be efficiently checked using a linear programming (LP) solver.

Recall that our goal is to calculate an upper bound of $K$-th moment.
Hence we naturally want to minimize the upper bound $\eta_K(l_{\init}, \vec{x}_{\init})$ calculated by a supermartingale (see Thm.~\ref{thm:uppboundHigher}).
We can achieve this goal by setting $\eta_K(l_{\init}, \vec{x}_{\init})$, a linear expression over $U$, to the objective function of the linear programming problem
and ask the LP solver to minimize it.
%


A natural question would about the converse of the implication $\text{(\ref{eq:1811101318})}\,\Rightarrow\,\text{(\ref{eq:1811101317})}$.
The following theorem answers the question to some extent.
\begin{mytheorem}[affine form of Farkas' lemma (see e.g.~{\cite[Cor.~7.1h]{Schrijver86TLIP}})]
Let $A \in \real^{n \times m}$, $b \in \real^m$, $c \in \real^n$ and $d \in \real$.
If $\{ x \mid A x \le b \}$ is not empty, the following two conditions are equivalent.
\begin{itemize}
\item $\forall x \in \real^n, A x \le b \implies \transpose{c} x \le d$
\item $\exists y \in \real^m, \transpose{A} y = c \land \transpose{b} y \le d$
\qed
\end{itemize}
\end{mytheorem}

We note that
if a pCFG $\Gamma$ has no program variable ($V = \emptyset$) and all the transitions are probabilistic (that is, if $\Gamma$ is a Markov chain), the above method gives the exact value of moments. It is because the LP problem has the following obvious optimal solution: $\eta_k(l) = \text{(the $k$-th moment of runtimes from location $l$)}$.

%

\subsection{Polynomial Template-based Algorithm}
We consider fixing a polynomial template for a supermartingale.
The algorithm in this section is based on~\cite{DBLP:conf/cav/ChatterjeeFG16}.

\begin{mydefinition}\label{def:polyexp}
Let $V=\{x_1,\ldots,x_n\}$ be a set of variables. 
A \emph{monomial} is a formula of a form $x_{i_1}^{d_1}\dots x_{i_p}^{d_p}$.
We call $d_1+\cdots+d_p$ a \emph{degree} of the monomial, and 
write $\monomials{d}$ for the set of monomials whose degrees are no greater than $d$.
A \emph{polynomial expression} (or simply a \emph{polynomial}) over $V$ is a formula of a form 
$a_1m_1+\cdots+a_qm_q+b$ where $a_1,\ldots,a_q,b\in\real$ and $m_1,\ldots,m_q$ are monomials.
We write $\polyexpr{\real}{V}$ for the set of polynomial expressions over $V$.
The notions of \emph{polynomial inequality}, \emph{polynomial conjunctive predicate},
\emph{polynomial predicate}, \emph{polynomial expression map} and \emph{polynomial predicate map}
are defined in a similar manner to Def.~\ref{def:linexp}.
\end{mydefinition}

In the polynomial case, we assume that a pCFG $\Gamma$,
an invariant $I$, a set $C \subseteq I$ of configurations and a natural number $K$ satisfy the following conditions.

\begin{myassumption}\mbox{}
\begin{itemize}
\item For any $l \in L_A$, $\mathrm{Up}(l) = (x_j, u)$,
\begin{itemize}
\item if $u \in \Borel(\real^V, \real)$, then $u$ is represented by a polynomial expression over $V$
\item if $u \in \mathcal{D}(\real)$, the $K$-th moment of $u$ is known.
\item if $u \in \Borel(\real)$, then $u$ is represented by a polynomial predicate $\phi$ over $\{ x_j \}$.
\end{itemize}
\item For any $l \in L_D$, $l'$, $G(l, l') = \llbracket p \rrbracket$ is represented by a polynomial predicate $p$ over $V$.
\item the invariant $I$ is represented by a polynomial predicate map over $\Gamma$.
\item the set $C$ of terminal configurations is represented by a polynomial conjunctive predicate map.
\end{itemize}
\end{myassumption}

The polynomial template-based synthesis algorithm is similar to the linear template-based one. 
In the polynomial case, the user have to fix the maximum degree $d$ of the polynomial template.
The algorithm first fixes a $d$-degree polynomial template for a supermartingale.
It is a family of formulas indexed by $i\in\{1,\ldots,K\}$ and $l\in L$ that have the following form:
\[
\eta_i(l, \vec{x}) \;=\;
\sum_{m\in\monomials{d}}
a^l_{m,i}m+ b^l_{i}\,.
\]
Each $a^l_{m,i}$ and $b^l_{i}$ are newly added variables called \emph{parameters},
and we write $U$ for the set of all parameters. 
Our goal is to find a valuation $U\to\mathbb{R}$ so that a $K$-tuple
$\bigl(\eta_1(\place, \vec{x}),\ldots,\eta_K(\place, \vec{x})\bigr)$
of polynomial expression maps is
a ranking supermartingale for $K$-th moment (Definition~\ref{def:ranksupHigher}).

Similarly to the linear case, the algorithm collects conditions on the parameters. 
It results in a conjunction of formulas of the following form:
\begin{equation}\label{eq:1811101315poly}
\forall \vec{x}\in\real^V.\; \varphi_1\rhd_1 0\wedge\cdots \varphi_m\rhd_m 0 \implies \psi\geq 0\,.
\end{equation}
Here $\rhd_i\in \{\geq,>\}$, each $\varphi_i$ is a polynomial expression without parameters, and 
$\psi$ is a polynomial formula over $V$ whose coefficients are \emph{linear} expressions over $U$.
Relaxing the strict inequalities, we obtain the following:
\begin{equation}\label{eq:1811101316poly}
\forall \vec{x}\in\real^V.\; \varphi_1\geq 0\wedge\cdots \varphi_m\geq 0 \implies \psi\geq 0\,.
\end{equation}

To reduce (\ref{eq:1811101316poly}) to a form that is solvable by a numerical method,
we can use the notion of \emph{sum-of-square polynomials}~\cite{DBLP:conf/cav/ChatterjeeFG16}. 
A polynomial expression $f$ is said to be \emph{sum-of-square} (SOS) if there exist polynomial expressions $g_1,\ldots,g_l$ such that 
$f=g_1^2+\cdots+g_l^2$. 

Obviously, a sum-of-square polynomial is nonnegative. 
Therefore the following formula is a sufficient condition for (\ref{eq:1811101316poly}):
\begin{equation}\label{eq:1811101317poly}
\exists \bigl(h_w:\text{sum-of-square}\bigr)_{w\in\{0,1\}^m}.\; \psi=\sum_{w\in\{0,1\}^m} h_w\cdot\varphi^{w_1}_1\cdot\dots\cdot \varphi^{w_m}_m\,.
\end{equation}
Here $w_i$ denotes the $i$-th component of $w\in\{0,1\}^m$.

One of the reasons that sum-of-square is convenient is that it is characterized using the notion of \emph{positive semidefinite matrix}.
\begin{myproposition}[see e.g.~\cite{Horn:2012:MA:2422911}]
A polynomial expression $f$ over $V$ is sum-of-square if and only if there exist a vector $\vec{y}$ whose components are monomials over $V$ and 
a positive semidefinite matrix $A$ such that $f=\transpose{\vec{y}}A\vec{y}$.
\qed
\end{myproposition}

By the proposition above, existence of a valuation $U\to\mathbb{R}$ of parameters and sum-of-square polynomials 
as in (\ref{eq:1811101317poly}) can be reduced to a \emph{semidefinite programming} (SDP) problem.
Likewise the linear case, by setting a linear expression $\eta_K(l_{\init}, \vec{x}_{\init})$ to the objective function,
we can minimize it.

In the linear case, completeness was partially ensured by Farkas' lemma.
In the polynomial case, the role is played by the following theorem.

\begin{mytheorem}[Schm\"udgen's Positivstellensatz~\cite{Schmudgen1991}]
Let $g, g_1, \dots, g_m$ be polynomial expression over a set of variable $V$. If $\{\vec{x}\in\real^V\mid \bigwedge_{i=1}^m g_i \ge 0 \}$ is compact, then 
the following conditions are equivalent:
\begin{itemize}
\item $\forall x \in \real^V.\; g_1 \ge 0\wedge\cdots\wedge g_m\ge 0 \implies g > 0$.
\item there exists a family $\{ h_w \}_{w \in \{0, 1\}^m}$ of sum-of-square polynomial expressions such that $g = \sum_{w \in \{0, 1\}^m} h_w \cdot g_1^{w_1}\cdot\cdots\cdot g_m^{w_m}$.
\qed
\end{itemize}
\end{mytheorem}


\section{Test Programs}\label{sec:testprogs}
We have augmented the standard syntax of randomized program (see e.g.~\cite{DBLP:conf/popl/ChatterjeeNZ17}) so that
we can specify an invariant and a terminal configuration.
To specify an invariant, we can use either of the following syntax.
\begin{itemize}
\item $\$\ldots$ specifies an invariant globally.
\item $\{\ldots\}$ specifies an invariant locally.
\end{itemize}
We can specify a terminal configuration by using $\mathsf{refute}(\ldots)$.
\begin{lstlisting}[numbers=left, frame=lines, basicstyle={\ttfamily\scriptsize}, caption={(1-1) \texttt{coupon\_collector}}]
$ 0 <= c0 and c0 <= 1 and 0 <= c1 and c1 <= 1

c0 := 0;
c1 := 0;

while true do
	if prob(0.5) then
		c0 := 1
	else
		c1 := 1
	fi;
	refute (c0 + c1 > 1)
od
\end{lstlisting}
\begin{lstlisting}[numbers=left, frame=lines, basicstyle={\ttfamily\scriptsize}, caption={(1-2) \texttt{coupon\_collector4}}]
$ 0 <= c0 and c0 <= 1 and 0 <= c1 and c1 <= 1 and 0 <= c2 and c2 <= 1 and 0 <= c3 and c3 <= 1

c0 := 0;
c1 := 0;
c2 := 0;
c3 := 0;

while true do
	if prob(0.5) then
		if prob(0.5) then
			c0 := 1
		else
			c1 := 1
		fi
	else
		if prob(0.5) then
			c2 := 1
		else
			c3 := 1
		fi
	fi;
	refute (c0 + c1 + c2 + c3 > 3)
od
\end{lstlisting}
\begin{lstlisting}[numbers=left, frame=lines, basicstyle={\ttfamily\scriptsize}, caption={(2-1) \texttt{random\_walk\_1d\_intvalued}}]
{ true } x := 1;

{x >= 1} while true do
{x >= 1}   if prob(0.6) then
{x >= 1}     x := x - 1
           else
{x >= 1}     x := x + 1
           fi;
{x >= 0}   refute (x < 1)
         od
\end{lstlisting}
\begin{lstlisting}[numbers=left, frame=lines, basicstyle={\ttfamily\scriptsize}, caption={(2-2) \texttt{random\_walk\_1d\_realvalued}}]
{ true }                          x := 2;

{ x >= 0 }                        while true do
{ x >= 0 }                          if prob(0.7) then
{ x >= 0 }                            z := Unif(0,1);
{ x >= 0 and 0 <= z and z <= 1 }      x := x - z
                                    else
{ x >= 0 }                            z := Unif(0,1);
{ x >= 0 and 0 <= z and z <= 1 }      x := x + z
                                    fi;
{ x >= -1 }                         refute (x < 0)
                                  od
\end{lstlisting}
\begin{lstlisting}[numbers=left, frame=lines, basicstyle={\ttfamily\scriptsize}, caption={(2-3) \texttt{random\_walk\_1d\_adversary}}]
{ true }               x := 2;

{ 0 <= x and x <= 13 } while true do
{ 0 <= x and x <= 10 }   if prob(0.8) then
{ 0 <= x and x <= 10 }     skip
                         else
{ 0 <= x and x <= 10 }     if prob(0.5) then
{ 0 <= x and x <= 10 }       x := x + 1
                           else
{ 0 <= x and x <= 10 }       x := x + 2
                           fi
                         fi;
{ 0 <= x and x <= 12 }   if * then
{ 0 <= x and x <= 12 }     if prob(0.875) then
{ 0 <= x and x <= 12 }       x := x - 1
                           else
{ 0 <= x and x <= 12 }       skip
                           fi
                         else
{ 0 <= x and x <= 12 }     if prob(0.8) then
{ 0 <= x and x <= 12 }       skip
                           else
{ 0 <= x and x <= 12 }       if prob(0.5) then
{ 0 <= x and x <= 12 }         x := x + 1
                             else
{ 0 <= x and x <= 12 }         x := x + 2
                             fi
                           fi;
{ 0 <= x and x <= 14 }     x := x - 1
                         fi;
{ 0 <= x and x <= 13 }   refute (x <= 0)
                       od
\end{lstlisting}
\begin{lstlisting}[numbers=left, frame=lines, basicstyle={\ttfamily\scriptsize}, caption={(2-4) \texttt{random\_walk\_2d\_demonic}}]
{ true }                                      x := 2;
{ x = 2 }                                     y := 2;
{ 0 <= x and 0 <= y }                         while true do
{ 0 <= x and 0 <= y }                           if * then
{ 0 <= x and 0 <= y }                             z := Unif (-2,1);
{ 0 <= x and 0 <= y and -2 <= z and z <= 1 }      x := x + z
                                                else
{ 0 <= x and 0 <= y }                             z := Unif (-2,1);
{ 0 <= x and 0 <= y and -2 <= z and z <= 1 }      y := y + z
                                                fi;
{ -2 <= x and -2 <= y }                         refute (x <= 0);
{ 0 <= x and -2 <= y }                          refute (y <= 0)
                                              od
\end{lstlisting}
\begin{lstlisting}[numbers=left, frame=lines, basicstyle={\ttfamily\scriptsize}, caption={(2-5) \texttt{random\_walk\_2d\_variant}}]
{ true }                           x := 3;
{ x = 3 }                          y := 2;
{ x >= y }                         while true do
{ x >= y }                           if * then
{ x >= y }                             if prob(0.7) then
{ x >= y }                               z := Unif (-2,1);
{ x >= y and -2 <= z and z <= 1 }        x := x + z
                                       else
{ x >= y }                               z := Unif (-2,1);
{ x >= y and -2 <= z and z <= 1 }        y := y + z
                                       fi
                                     else
{ x >= y }                             if prob(0.7) then
{ x >= y }                               z := Unif (-1,2);
{ x >= y and -1 <= z and z <= 2 }        y := y + z
                                       else
{ x >= y }                               z := Unif (-1,2);
{ x >= y and -1 <= z and z <= 2 }        x := x + z
                                       fi
                                     fi;
{ x >= y + 2 }                       refute (x <= y)
                                   od
\end{lstlisting}

\section{Detailed Comparison with Existing Work}
\subsection{Comparison with~\cite{DBLP:journals/corr/ChatterjeeF17}}
\label{appendix:comparisonWithChatterjeeF17}
In the literature on martingale-based methods, the one closest to this work is~\cite{DBLP:journals/corr/ChatterjeeF17}. Among its contribution is the analysis of tail probabilities by either of the following two combinations:
\begin{itemize}
 \item \emph{difference-bounded} ranking supermartingales and the corresponding choice of concentration inequality (namely Azuma's martingale concentration lemma); and
 \item (not necessarily difference-bounded) ranking supermartingales and Markov's concentration inequality.
\end{itemize}

 While implementation and experiments are lacking in~\cite{DBLP:journals/corr/ChatterjeeF17}, we can make the following theoretical comparison between these two methods and ours. 
\begin{itemize}
 \item The first method (with difference-bounded supermartingales) requires trying many difference bounds $c$, synthesizing a martingale for each $c$,  and picking the best one. This ``try many and pick the best''  workflow is much like in~\cite{DBLP:conf/popl/ChatterjeeNZ17}; it increases the computational cost, especially in the case a polynomial template is used (where a single synthesis procedure takes tens of seconds). 
 \item The second method corresponds precisely to the special case of our method where we restrict to the first moment. We argued that using higher moments is crucial in obtaining tighter bounds as the deadline becomes large, theoretically (\S\ref{sec:concentration_inequality}) and experimentally (\S{}\ref{sec:experiments}). 
\end{itemize}

\subsection{Comparison with~\cite{DBLP:conf/qest/KaminskiKM16}}
\label{appendix:comparisonWithKaminskiKM16}
In the predicate-transformer approach, the work~\cite{DBLP:conf/qest/KaminskiKM16}
 is the closest to ours, in that it studies \emph{variance} of runtimes of randomized programs. The main differences are as follows: 1) computing tail probabilities is not pursued; 2) their extension from mean to variance involves an additional variable $\tau$, which poses a challenge in automated synthesis as well as in generalization to even higher moments; and 3) they do not pursue automated analysis.

Let us elaborate on the above point 2). 
In syntax-based static approaches to estimating variances
		     or second moments, it is inevitable to simultaneously reason about both first and second moments. See Def.~\ref{def:timeElapseFunc}. 
		     In this work, we do so systematically by extending a notion of
		     supermartingale into a \emph{vector-valued} notion; this way our
		     theory generalizes to moments higher than the second in a
		     straight-forward manner. In contrast,
		     in~\cite{DBLP:conf/qest/KaminskiKM16},  an additional variable $\tau$---which stands for the
		     elapsed time---is used for mixing first and second
		     moments.

Besides the problem of generalizing to higher moments, a main drawback of this approach in~\cite{DBLP:conf/qest/KaminskiKM16} is that the
		     degrees of templates become bigger when it comes to automated synthesis. For example, due to
		     the use of $\tau^{2}$ in the condition for $\hat{X}$
		     in~\cite[Thm.~7]{DBLP:conf/qest/KaminskiKM16}, if the
		     template for $\tau$ is of degree $k$, the template for
		     $\hat{X}$ is necessarily of degree $2k$ or higher. One
		     consequence is that a fully LP-based implementation of the
		     approach of~\cite{DBLP:conf/qest/KaminskiKM16} becomes
		     hard, while it is possible in the current work (see~\S{}\ref{sec:experiments}). 

 Let us also note
		     that the work~\cite{DBLP:conf/qest/KaminskiKM16} focuses on
		     precondition calculi and does not discuss automated
		     synthesis or analysis.

\section{An Example of Polynomially Decreasing Tail Probability}\label{appendix:exTailProb}
We show that there exists a randomized program such that the tail probability of the runtime is polynomially decreasing (not exponentially decreasing).
A similar example can be found in~\cite[Example 8]{DBLP:journals/corr/ChatterjeeF17}.

\begin{lstlisting}[numbers=left,basicstyle={\scriptsize\ttfamily},frame=lines]
$ 0 <= r and r <= 1 and 0 <= n
n := 1;
r := Unif(0, 1);
while r * (n + 1) * (n + 1) <= n * n do
    r := Unif(0, 1);
    n := n + 1
od
\end{lstlisting}
Let $T_l$ be a random variable that represents the number of iterations.
As the program executes the loop body with probability $\frac{n^2}{(n+1)^2}$ in the $n$-th iteration, the tail probability of the runtime of the program is polynomially decreasing:
\[ \prob{T_l \ge d} = \left( \frac{1}{2} \right)^2 \cdots \left( \frac{d}{d+1} \right)^2 = \left( \frac{1}{d+1} \right)^2. \]

We can apply the polynomial template-based algorithm for this program (but cannot apply the linear one since the condition in the while statement is not linear).
Our implementation gives the following upper bound of the first moment of the runtime.
This upper bound can be used to bound tail probabilities, via the inequality in Prop.~\ref{prop:tail_ineq}.
\begin{center}
\begin{tabular}{|c|c|c|c|}
\hline
moment & upper bound & time (sec) & degree \\
\hline
1st & 13.15 & 534.575 & 3 \\
\hline
\end{tabular}
\end{center}

\end{document}